\documentclass{iopart}
\usepackage[utf8]{inputenc}
\usepackage{iopams}
\usepackage{geometry}
\usepackage{graphicx}
\usepackage{float}
\usepackage{subfig}
\usepackage[breaklinks=true,colorlinks=true,linkcolor=blue,urlcolor=blue,citecolor=blue]{hyperref}

\graphicspath{ {./feature_selection/}{./ml_tools/} {./tsne/}{./conf_mat/}{./layer_config/}{./proba_tsne/}{./env_ch/}{./later/}}

\begin{document}	
	\title{New methods to assess and improve LIGO detector duty cycle}
	
	\author{%
		A~Biswas$^{1,2,6}$, 
		J~McIver$^{2,3}$, 
		and
		A~Mahabal$^{2,4,5}$
	}%
	\medskip
	\address {$^{1}$IIT Gandhinagar, Gandhinagar, Gujrat 382355, India }
	\address {$^{2}$LIGO Laboratory, California Institute of Technology, Pasadena, CA 91125, USA }
	\address {$^{3}$University of British Columbia, Vancouver, Canada }
	\address {$^{4}$Division of Physics, Mathematics, and Astronomy, California Institute of Technology, Pasadena, CA 91125, USA }
	\address {$^{5}$Center for Data Driven Discovery, California Institute of Technology, Pasadena, CA 91125, USA }
	\address{$^{6}$University of California San Diego, 9500 Gilman Dr, La Jolla, CA 92093, USA}
	
	\begin{abstract}
		A network of three or more gravitational wave detectors simultaneously taking data is required to generate a well-localized sky map for gravitational wave sources, such as GW170817. 
		Local seismic disturbances often cause the LIGO and Virgo detectors to lose light resonance in one or more of their component optic cavities, and the affected detector is unable to take data until resonance is recovered. 
		In this paper, we use machine learning techniques to gain insight into the predictive behavior of the LIGO detector optic cavities during the second LIGO-Virgo observing run.
		We identify a minimal set of optic cavity control signals and data features which capture interferometer behavior leading to a loss of light resonance, or \textit{lockloss}.
		We use these channels to accurately distinguish between lockloss events and quiet interferometer operating times via both supervised and unsupervised machine learning methods. 
		This analysis yields new insights into how components of the LIGO detectors contribute to lockloss events, which could inform detector commissioning efforts to mitigate the associated loss of uptime.
		Particularly, we find that the state of the component optical cavities is a better predictor of loss of lock than ground motion trends.
		We report prediction accuracies of 98\% for times just prior to lock loss, and 90\% for times up to 30 seconds prior to lockloss, which shows promise for this method to be applied in near-real time to trigger preventative detector state changes.
		This method can be extended to target other auxiliary subsystems or times of interest, such as transient noise or loss in detector sensitivity. 
		Application of these techniques during the third LIGO-Virgo observing run and beyond would maximize the potential of the global detector network for multi-messenger astronomy with gravitational waves. 
	\end{abstract}
	
	\section{Introduction}\label{intro}
	
	Gravitational waves, small ripples in the fabric of spacetime, are able to probe the inner dynamics of highly energetic systems that are difficult to directly observe via electromagnetic radiation. The LIGO and Virgo collaborations have reported the detection of gravitational waves from the merger of ten binary black hole (BBH) systems~\cite{O2catalog}, and two binary neutron star (BNS) mergers~\cite{GW170817,GW190425} from their first, second, and third observing runs thus far. Four of these signals were identified in and analyzed across a global network of three gravitational wave interferometers: two LIGO detectors in the U.S. and Virgo in Italy \cite{aLIGO, adVirgo}. This global network enabled sky localization for three BBH signals (GW170814, GW170817, GW170818) on the order of tens of square degrees, with the BNS event GW170817 localized to just 16 square degrees~\cite{O2catalog}. The skymap produced for GW170817 resulted in the identification of the corresponding host galaxy \cite{Coulter:18, MMA}, which was the first example of multi-messenger astronomy with gravitational waves, and resulted in a broad array of advances in astrophysics \cite{Hubble, GammaRay}. 
	
	Maximizing observatory uptime is critical to achieving a high rate of confident detections and producing an accurate skymap with the broadest array of detectors in the global network.  
	The Advanced LIGO and Advanced Virgo detectors are currently in the midst of their third observing run (O3). Detector sensitivity is significantly higher than during O2; LIGO-Livingston has achieved an average sensitivity of over 130 Mpc, LIGO-Hanford of over 100 Mpc, and Virgo of just under 50 Mpc~\cite{PublicSummaryPages}. KAGRA is expected to join O3 toward the end of the observing run. The combined network was expected to detect as many as one significant GW event every week \cite{ObsScenario}, and the LIGO-Virgo Collaboration has released about one un-retracted candidate event per week via public alerts on average since the start of the run.  This three detector network has reported sky localizations for candidate events as low as tens of square degrees during O3~\cite{GraceDB}.
	However, ground-based gravitational wave detectors such as LIGO and Virgo are susceptible to loss in observation time, often due to elevated ground motion, including earthquakes, winds, microseism and anthropogenic activity. 
	
	Prior studies have shown a direct correlation between local  ground velocity and \textit{lockloss} events, when one or more optical cavities lose light resonance~\cite{Biscans, Mukund-19}. 
	For example, control of the component optic cavities within the detector becomes unstable at higher ground velocities.
	It can take several hours for a detector to return to its operating state after an earthquake, and high winds or high microseism may also prevent a stable detector control state.
	The LIGO detectors also record the time-series of power circulating within optic cavities and time-series used to monitor and control cavity length and the relative angle of the component optics. 
	By using information recorded in these \textit{auxiliary channels}, we can infer which components of a detector are more susceptible to losing light resonance and causing a lockloss. Additionally, an accurate prediction of a likely lockloss and its cause would enable the detectors to engage a ``lockloss robust" state, which could allow the detector to resume nominal operation much more quickly than the minimal tens of minutes and often hours spent recovering from a lockloss event. 
	
	In recent years, machine learning has been employed to characterize noise in the LIGO and Virgo detectors, including the identification of glitches~\cite{Powell:2015, Powell:2017, Mukund:2017, George:2018}, the correlation of glitches with auxiliary witness sensors~\cite{Biswas:2013,Cavaglia:2019,Essick:2019}, and a novel approach that blends citizen science with convolutional neural nets~\cite{Zevin:2017, Coughlin:2019}.
	Deep learning methods in particular have been instrumental in characterizing rich structure in data that has lead to new insights into instrument behavior~\cite{Mukund:2019}. 
	Deep learning methods internally explore millions of features and internally downselect a subset that provide the best discrimination~\cite{deep_learning}. This is in contrast to tens to hundreds of features that humans might handpick. While these handpicked domain knowledge assisted features are often a great start, they also tend to be biased and almost always incomplete. Deep learning supplements these with complex features directly derived from the data. 
	In this study we explore deep learning techniques to understand complex nonlinear interaction of auxiliary channels preceding lockloss events, with the goal of improving the duty cycle of the global interferometer network.\par 
	
	In this paper we present results using O2 data to illustrate the effectiveness of three different machine learning approaches with different strengths to improve our understanding of why the Advanced LIGO interferometers lose lock. In Section 2 we describe the data we used, including the selection of labeled lockloss and `quiet' times. In Section 3 we detail the pre-processing of the data prior to the application of machine learning tools. In Section 4 we summarize the machine learning tools we applied and the feature sets we used for ML each approach. In Section 5 we present our results. In Section 6 we discuss our results and outline future targets for this effort as the LIGO and Virgo detectors prepare for their fourth observing run.

	
	\section{Data}\label{sec:data} 
	
	\subsection{Auxiliary channel data}\label{Auxiliary channel data}
    Each LIGO detector records data from roughly 200,000 auxiliary channels that witness the behavior of the detectors' subsystems and local environment~\cite{Effler:2015,Covas:2018}. 
	To limit the computational expense of our analysis, we downselected to 22 channels, including ground motion sensors that witness seismic events and channels used to sense and control the optics comprising the major resonant cavities in the interferometer. 
	
	For ground motion witnesses, we chose auxiliary channels that would form a minimal basis set to distinguish between different sources of seismic events. Based on the prior finding that LIGO detector lockloss mechanisms include ground motion \cite{Biscans}, we investigated the following potential sources of seismic-related locklosses: earthquakes, high wind, high microseismic ground motion, and anthropogenic noise in the 10-30 Hz band, which captures ground motion due to nearby human activity such as snow removal at the LIGO-Hanford site. For all ground motion witness channels, we used the array of Streckeisen STS-2s sensors at both LIGO sites~\cite{SEI,pemsite}. Since higher frequency ground motion tends to be highly localized and change rapidly, we included downsampled second-trends\footnote{For second-trends, we used data averaged over each second. Similarly, we used data averaged over each minute as minute-trends.} of channels that witnessed 10-30 Hz vertical ground motion at the two \textit{end stations} which house the optics used for one end of each of the interferometer arms, and the \textit{corner station}, which houses the input laser, beam splitter, auxiliary optics, and data readout. To distinguish between earthquake events, which manifest in vertical and horizontal ground motion at low frequencies, and wind, which induces ground tilt sensed by the seismometers as horizontal ground motion in the same frequency band, we included a second-trend of 0.03-0.1 Hz vertical ground motion and another witnessing horizontal ground motion in the same frequency band. To capture microseismic ground motion, we also included a downsampled minute-trend of 0.1-0.3 Hz vertical ground motion \cite{bp_env}.  
	
\begin{figure}
		\includegraphics[width=0.9\textwidth]{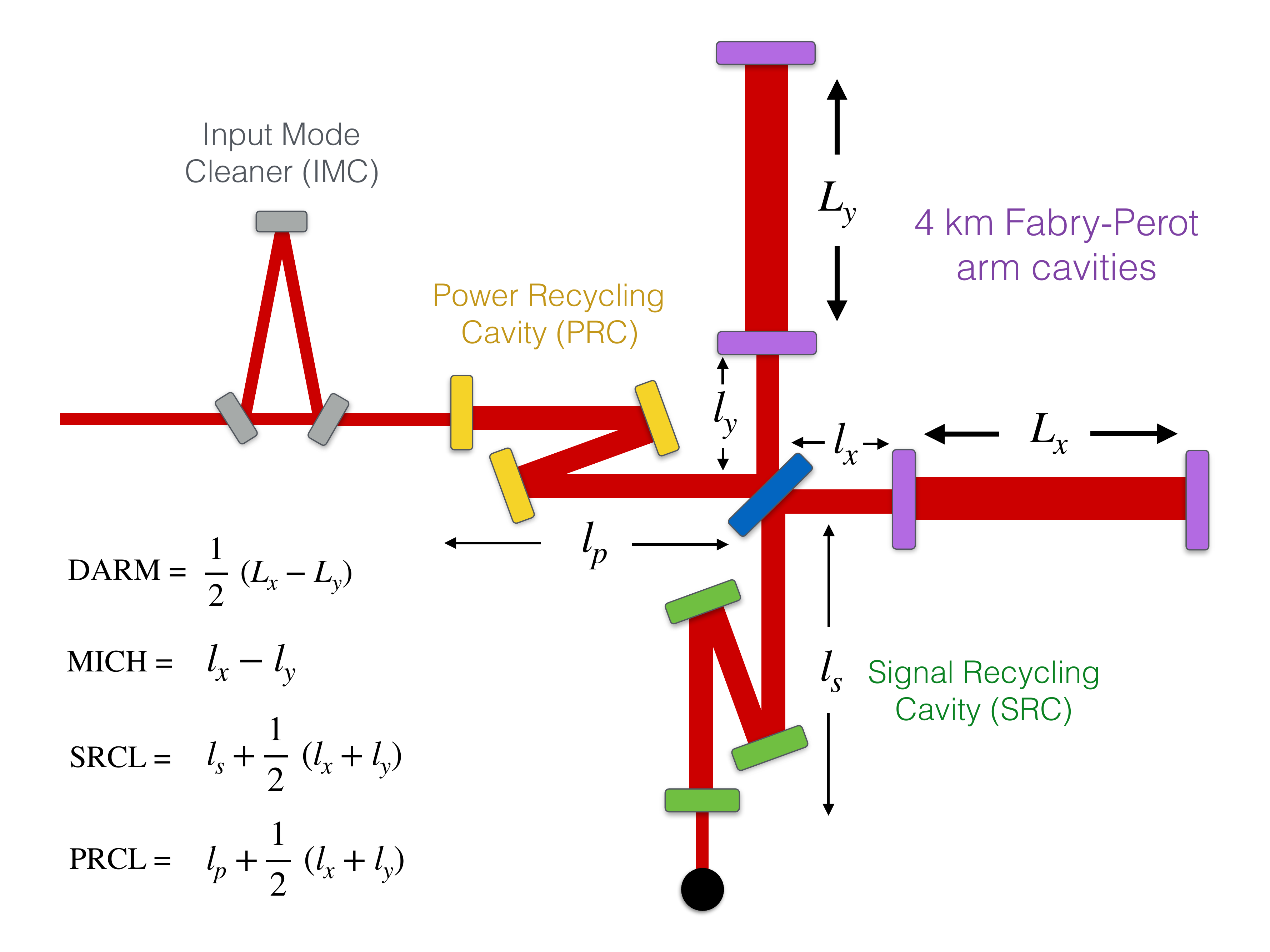}
		\centering
		\caption{Simplified schematic of the basic Advanced LIGO interferometer layout, illustrating how differential arm length (DARM), the inner Michelson interferometer (MICH), the Signal Recycling Cavity Length (SRCL) and Power Recycling Cavity Length (PRCL) are calculated relative to the position of the beam splitter optic, shown in blue.}\label{fig:ifo_layout}
	\end{figure}

	For interferometer optic cavity control channels, again based on prior knowledge of potential lockloss mechanisms, we chose a channel set that spanned different degrees of freedom used to actively stabilize cavities both in length and relative angle of the component optics. To target length sensing and control, we included a witness of power in the Input Mode Cleaner (IMC), power in the Power Recycling Cavity (POP), and light reflected from the Power Recycling Mirror (REFL). We also included length control signals of the inner Michelson interferometer (MICH), the Power Recycling Cavity (PRCL), and the Signal Recycling Cavity (SCRL), as shown in Figure \ref{fig:ifo_layout}. To target alignment sensing and control, we included control signals for the `pitch' and `yaw' degrees of freedom for the common and differential modes of the interferometer, and the `hard' and `soft' relative optic orientations of each (see Figure 16 of \cite{aLIGO}).  
	
	For each of the 22 auxiliary channels we studied, we considered time-series of a fixed duration before lockloss. We then extracted several features described in the next Section (e.g. {\it dmdt} mappings, spectrograms and various statistical features) from the time-series data. We applied a variety of machine learning techniques on these computed features to visualize differences in behavior as clusters and classified the inputs into two classes: lockloss or quiet (non-lockloss). 
	
	\subsection{Event selection}\label{event selection}
	For labelled lockloss times, we used the \textit{Guardian} state machine platform~\cite{Rollins2016} to identify 942 times when the LIGO-Hanford detector transitioned from a nominal low noise state to the lockloss state during O2. We excluded any lockloss event that occurred within 5 minutes of another lockloss event. For more precise timing information, we used pre-computed lockloss times corresponding to the time when the variance of an indicator channel, such as a photodiode, exceeded a standard deviation. 
	Additionally, we identified 1094 quiet times in observation mode which were at least 4 minutes away from lockloss events; a subset of the total number of possible quiet times during the run. 
	We reserved 25\% of lockloss and quiet time examples as a test set and we used the remaining examples for constructing the training set.
	
	We analyzed the minimum time duration prior to lockloss that captured the variance of the channel for different types of channels. For channels monitoring the sensing and control of the interferometer optic cavities, we used a duration of 30~seconds prior to lockloss. For 0.03-0.1~Hz ground motion, where we would expect earthquake and wind event to be most evident, we used a duration of one hour. For 0.1-0.3~Hz ground motion, where microseismic ground motion dominates, we used a duration of 24~hours. For 10-30~Hz ground motion, which witnesses ground motion induced by human activity, we used a duration of 10 minutes.


	\section{Data preparation}\label{data_prep}
	
	Our analysis methods, particularly convolutional neural networks, require representative images as input. Below we detail how we converted the time-series data to images. 
	
	\subsection{Optimizing {\it dmdt} features}\label{Deciding binning of dmdt features}
	
\begin{figure}
		\includegraphics[width=.9\textwidth]{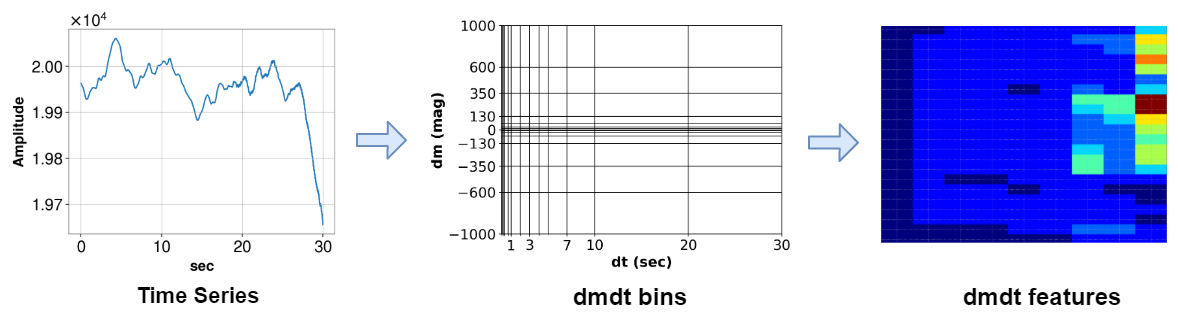}
		\centering
		\caption{Converting a time-series into a {\it dmdt} mapping as described in \ref{Deciding binning of dmdt features}. The left panel is the time-series, the middle panel shows the {\it dmdt} bins, and the third is the {\it dmdt} embedding where unequal sized {\it dmdt} bins are converted to equal-sized pixels with the pixel values assigned as per bin occupancy. As described in ~\cite{Richards2011}, for astronomy {\it dm} is the negative log of auxiliary channel time-series amplitude, but depending on the application it can be any quantity, logarithmic or otherwise. }
		\label{fig:dmdt}
	\end{figure}

	We considered image classification due to a similar successful application to astronomical light curves (time-series) \cite{light_curve}. 
	In this method, the time-series is converted to a 2D representation (called \textit{dmdt}) by considering all pairwise differences in time ({\it dt}) and the observable ({\it dm}), followed by the binning of 
	the differences over a range of defined {\it dm} and {\it dt} intervals. 
	The 2D plot so obtained is called a {\it dmdt}-image (see Figure.~\ref{fig:dmdt}).
	These images are then used to train a convolutional neural network (commonly known as a CNN or conv-net) for classification. 
	Our observed amplitudes have been treated equivalent to magnitudes in the astronomy case, and correspondingly we took exponentials of the amplitudes for calculating the percentile ratios for consistency. The method itself is agnostic to the use of linear, logarithmic and exponential values and will reject features that do not contribute to the overall discriminatory power in a given problem.
	
	In order to maximize the performance of the convolutional neural net, we optimized our chosen feature set to best capture the behavior exhibited by our selected auxiliary channels prior to loss of lock. 
	Here we outline our procedure for determining the best approach for conditioning the time-series data we projected onto the {\it dmdt} feature space, and deciding the binning ({\it dm} and {\it dt} bins) for the {\it dmdt} features. We used strong earthquake times (0.03-1~Hz ground motion velocity $>$ 500 nm/s) when we expected to see a resulting loss of lock in order to isolate frequency ranges where auxiliary channels exhibited a response to the elevated ground motion prior to the lockloss event. We repeated this procedure for 10 examples to establish a range of channel responses to a strong seismic event. 
	
\begin{figure}
		\centering
		\subfloat[]{\includegraphics[width=0.8\textwidth]{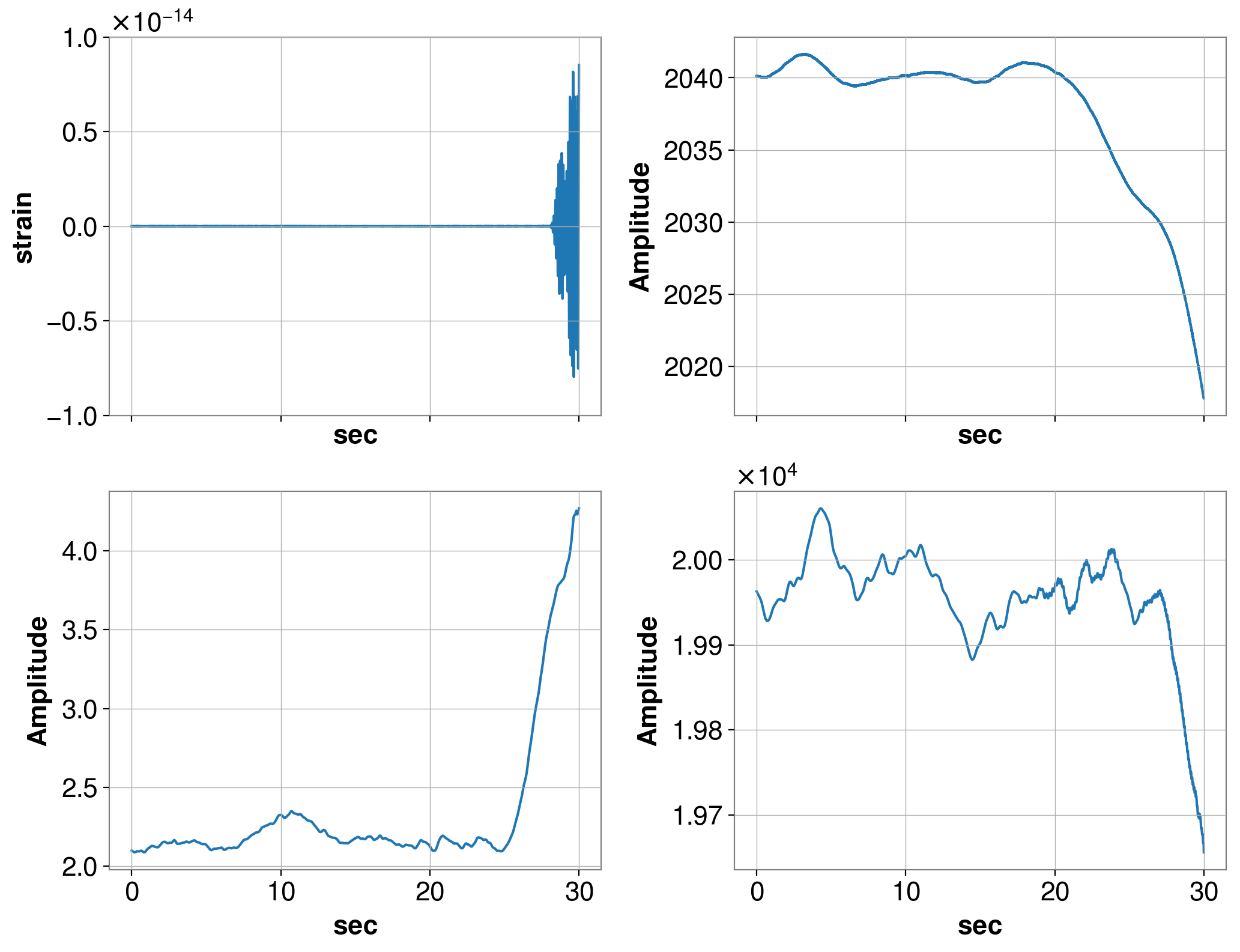}}
		\hfill
		\subfloat[]{\includegraphics[width=0.8\textwidth]{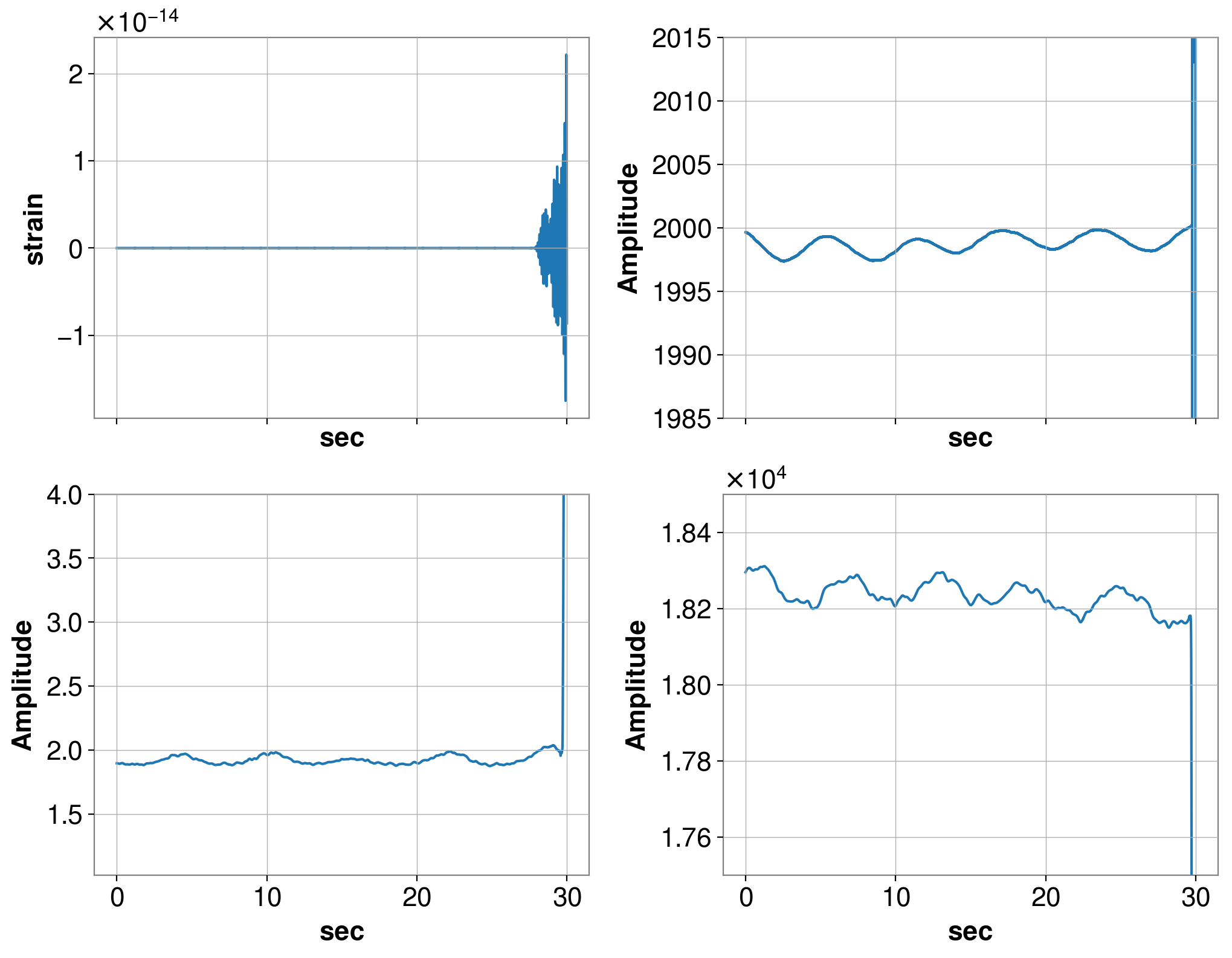}}
		\caption{For selection of interesting frequency ranges to use to construct distinguishing features, we used lockloss events where auxiliary channels show interesting behavior prior to the lockloss, as shown in (a) for the lockloss event (2017-01-10 06:36:56 UTC), as opposed to responding to the lockloss, as shown in (b) for the event (2017-01-19 23:11:06 UTC). In plots (a) and (b), gravitational-wave strain (top-left) and the auxiliary channels IMC (top-right), REFL (bottom-left) and POP (bottom-right) are shown and the recorded lockloss occurs at the end of each time-series.\label{fig:time_compare}}
	\end{figure}

	
	{\bf Down selecting interesting examples} For feature set tuning, we compared time-series of the analyzed auxiliary channels to the strain channel for each of the lockloss times corresponding to the 10 selected earthquakes. In some examples, we observed that there is some activity prior to behavior caused by a lockloss, as shown in the top series of plots in Figure \ref{fig:time_compare}. We considered such examples in the process of deciding the corner frequencies used for calculating band-limited root-mean-square trends, as described below. We did not consider examples where auxiliary channel behavior was responding to the loss of detector lock (instead of preceding it), as shown in the bottom series of plots in Figure \ref{fig:time_compare}. 
	
\begin{figure}
		\includegraphics[width=1\textwidth]{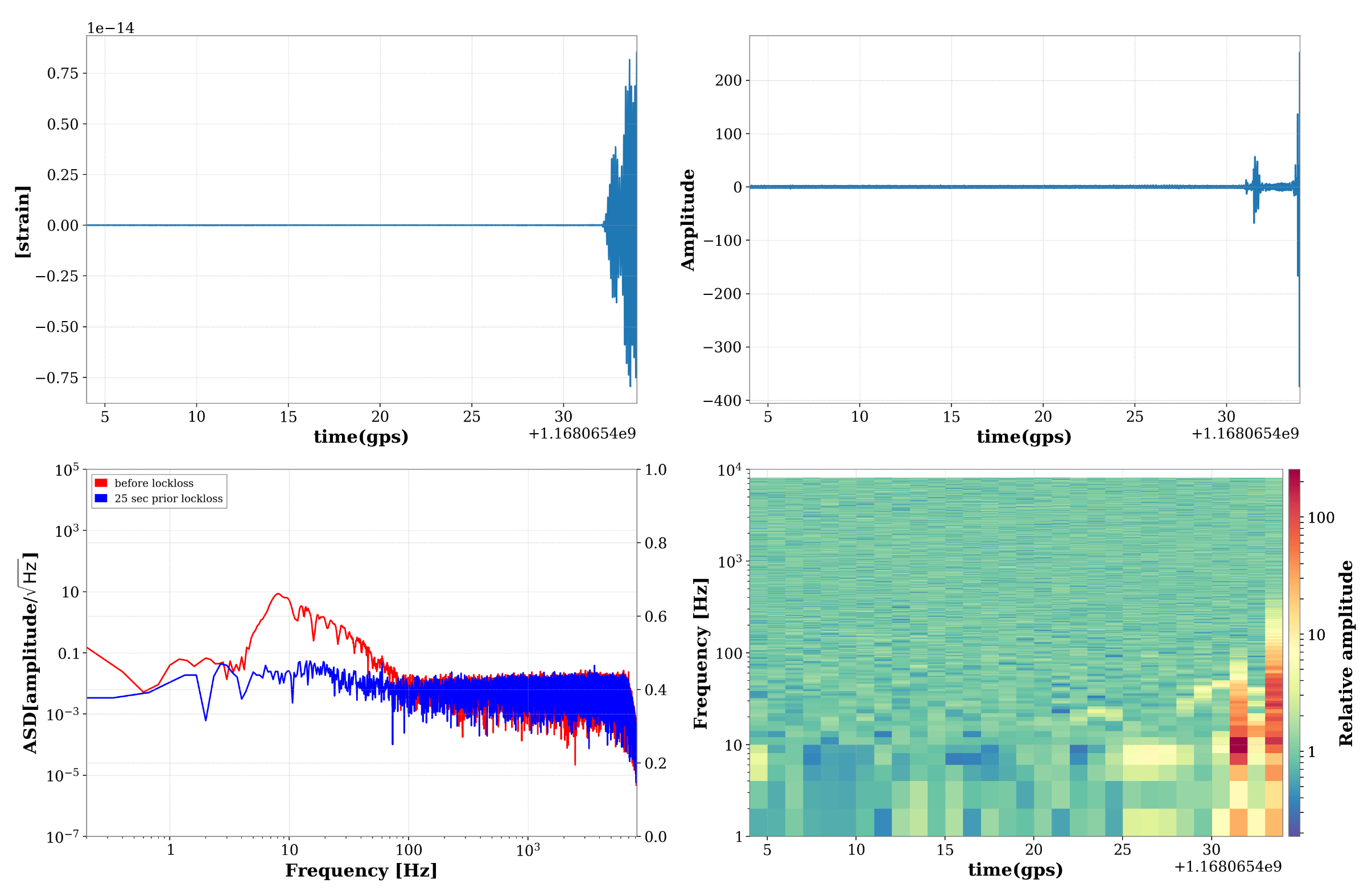}
		\centering
		\caption{An example of the visualization workflow used to decide the frequency range that best captured behavior prior to lockloss for each analyzed channel. The raw gravitational wave strain time-series just prior to a lockloss is shown in the top left plot, where the recorded lockloss occurs at the end of the time-series. A raw time-series of MICH using same time duration is shown in the top right plot.  Relative amplitude spectral densities and a median-normalized spectrogram of MICH are shown in the bottom left and right plots, respectively. }
		\label{fig:corner_freq}
	\end{figure}

	{\bf Deciding corner frequencies for bandpassed \textit{dmdt} features} 
	As a distinguishing feature set, \textit{dmdt} is often more effective when a bandpassed time-series highlighting only the frequencies that capture the behavior of interest is used as input instead of a raw time-series. 
	In order to determine the frequency band that best captured channel activity prior to lockloss, we evaluated the change in spectral content just prior to lockloss. We visualized this behavior using relative spectra comparing pre-lockloss times and quiet times as well as median-normalized time-frequency spectrograms, as shown in Figure \ref{fig:corner_freq}. 
	For the `pre-lockloss' spectrum we used 1-5 seconds prior to lockloss, excluding the second just prior to the recorded lockloss in order to avoid contaminating our results with behavior responding to loss of light resonance. 
	For the `quiet' spectrum we used 25-29 seconds prior to lockloss. 
	After selecting the corner frequencies for each of the analyzed channels, we bandpassed and down-sampled the data before passing them to \textit{dmdt}.

\begin{figure}
		\includegraphics[width=0.75\textwidth]{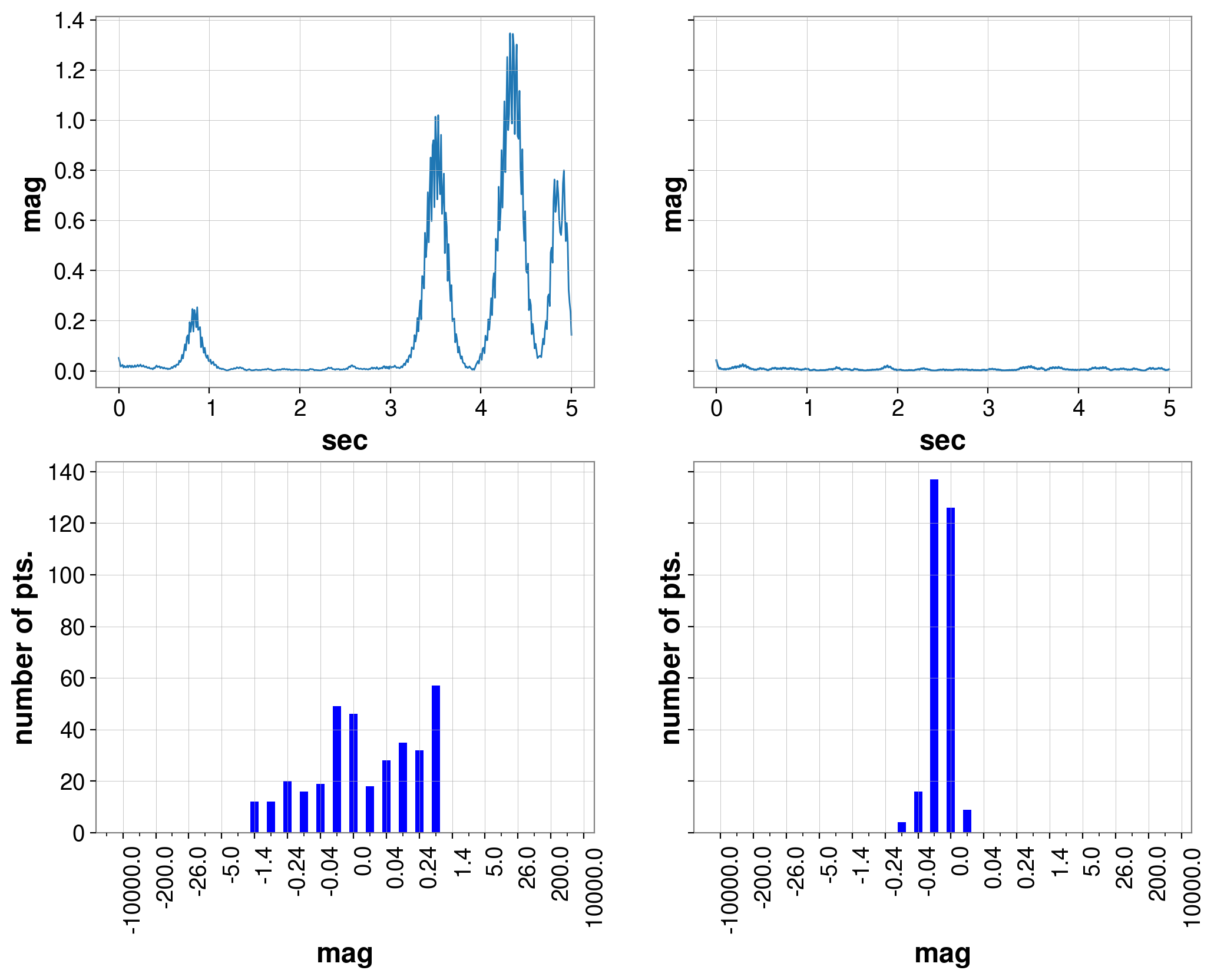}
		\centering
		\caption{An illustration of how our chosen {\it dm} bins capture a lockloss time relative to a quiet time. A lockloss example is shown on the left, and a quiet example on the right. The top plots show a bandpassed time-series of POP, and the bottom plots show \textit{dm} histograms of that POP data.} 
		\label{fig:dmhist}
	\end{figure}

	{\bf Deciding \textit{dm} and \textit{dt} bins}
	We chose {\it dt} such that each bin contained at least 10 samples. 
	We distributed {\it dm} bins for each channel such that the bins covered the range of differences in magnitude ({\it dm}). We ensured that most of the bins are filled while we are able to distinguish between the {\it dm} histogram plots of the lockloss and the quiet period. We selected different {\it dm} bins for each channel to reflect differences in the base time-series magnitude. Figure~\ref{fig:dmhist} shows an example of how dm bins capture a lockloss time compared to a quiet time.
	
	
	\subsection{Spectrograms}
\begin{figure}
		\centering
		
		\subfloat[]{\includegraphics[width=0.45\textwidth]{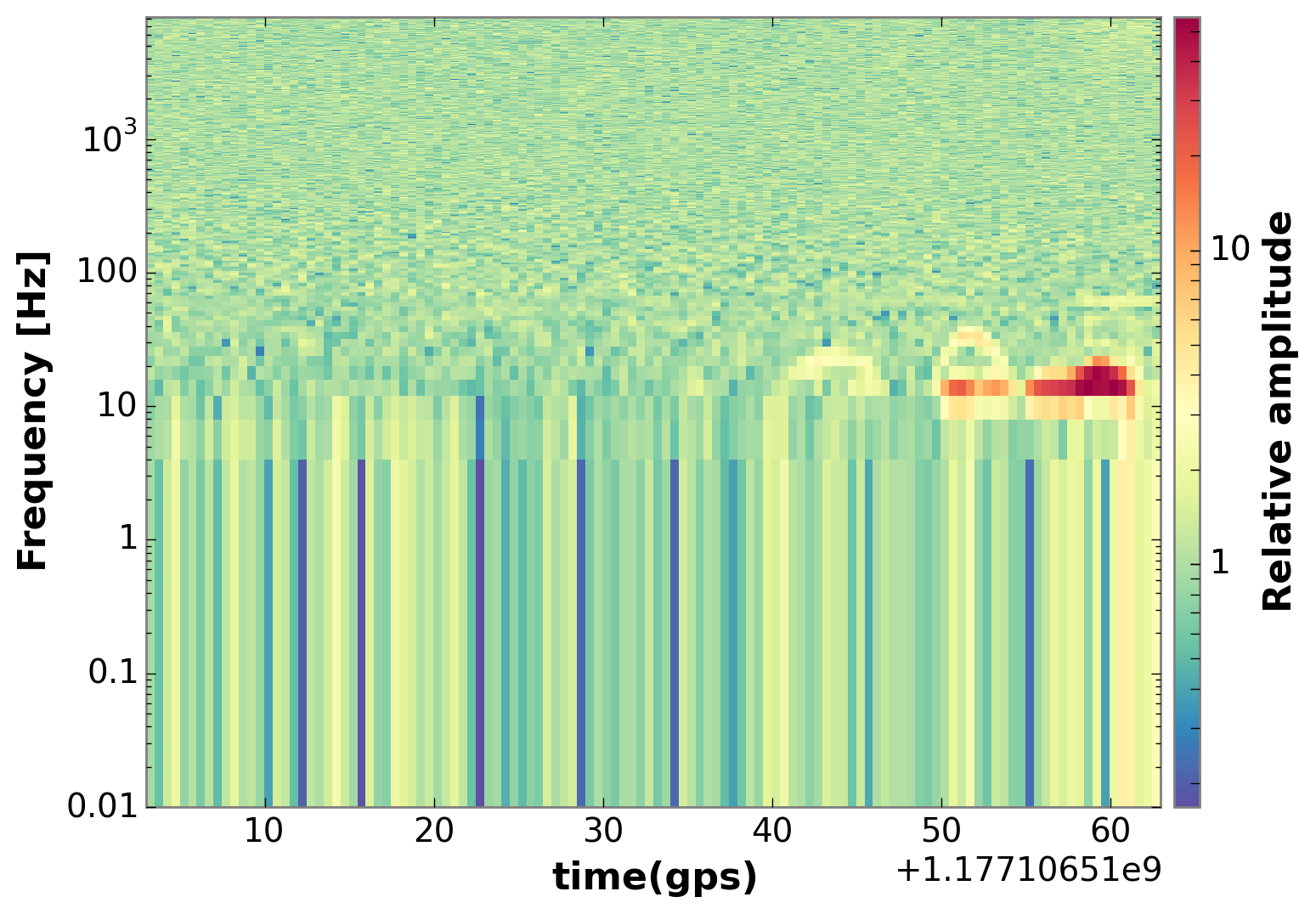}}
		\hfill
		\subfloat[]{\includegraphics[width=0.45\textwidth]{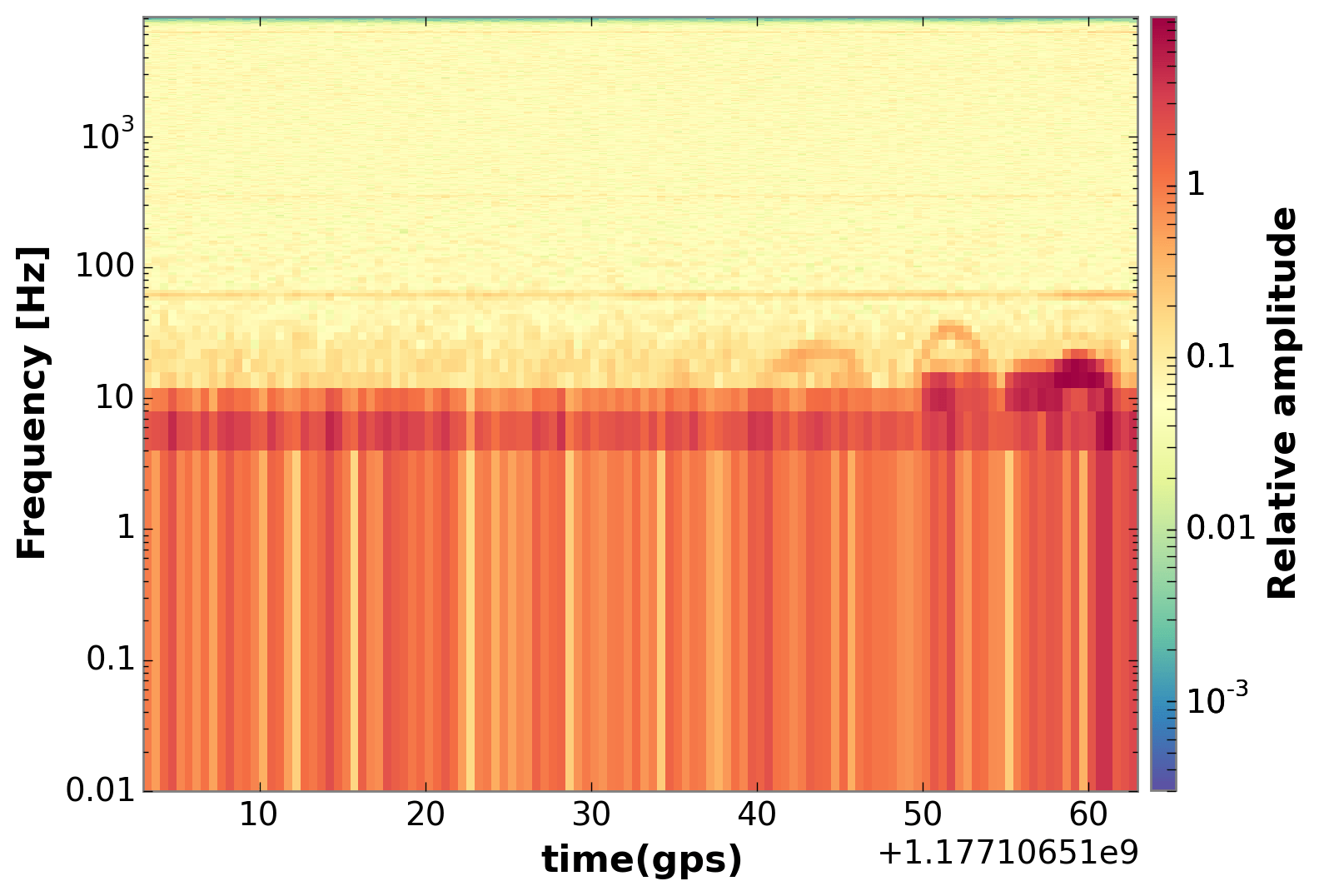}}
		\caption{Examples of  LIGO auxiliary channel spectrograms generated for CNN classification. Median-normalized spectrogram (left) and unnormalized spectrogram (right) of duration 30 seconds for SRCL channel. 	These spectrograms were produced using a time bin width of 0.25 seconds,  PSDs generated using an fft-length of 0.5 seconds, and overlap between PSD estimations of 50\%.}
		\label{fig:spectrogram}
	\end{figure}

		We also used spectrograms, another image representation of the time-series. Spectrograms are two-dimensional time-frequency decompositions of a time-series that represent the amplitude of the data over both time and frequency. 
	We consider both median-normalized spectrograms, where the data in each time and frequency bin normalized to the median amplitude for the corresponding frequency bin, and unnormalized spectrogram as features for a CNN. Examples of the two types of spectrograms and the parameters used to produce them are shown in Figure~\ref{fig:spectrogram}. Figure~\ref{fig:cnn} shows the workflow of transforming time-series to {\it dmdt} images and spectrograms for CNN classification.
	
	\section{Analysis methods}\label{analysismethods}
	\subsection{Machine learning tools used}
	\label{mltools}
We describe below each of the machine learned tools we targeted, the data features used, and the preprocessing applied.
Of the many machine learning techniques available, we chose to focus on three: (a) random forests (RF; stable and fast supervised learning \cite{rf}), (b) convolutional neural networks (CNNs; supervised deep learning \cite{deep_learning}), and (c) t-distributed stochastic neighbour embedding (t-SNE; unsupervised clustering \cite{tsne}). The high data sampling rates of the 22 selected LIGO auxiliary channels (up to 512 Hz) motivated additional data volume reduction through systematic pre-processing before passing the data to machine learning tools. 
	
	\subsubsection{Random forest}\label{Random forest}
	
\begin{figure}
		\includegraphics[width=0.9\textwidth]{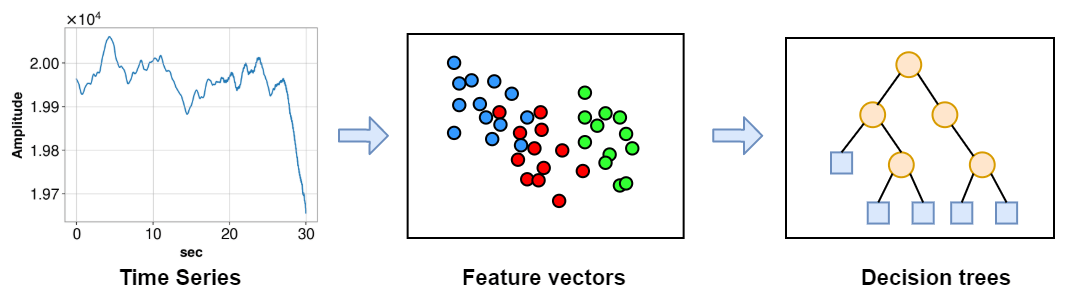}
		\centering
		\caption{Work-flow for classification with random forest. The left panel shows a time-series, the middle panel depicts features derived from the time-series (such as those described in Table~\ref{table:rf_features}), and the right panel shows a series of possible decision trees that determine the likely class of the source of the time-series.}\label{fig:random forest}
	\end{figure}

\begin{figure}
		\includegraphics[width=0.9\textwidth]{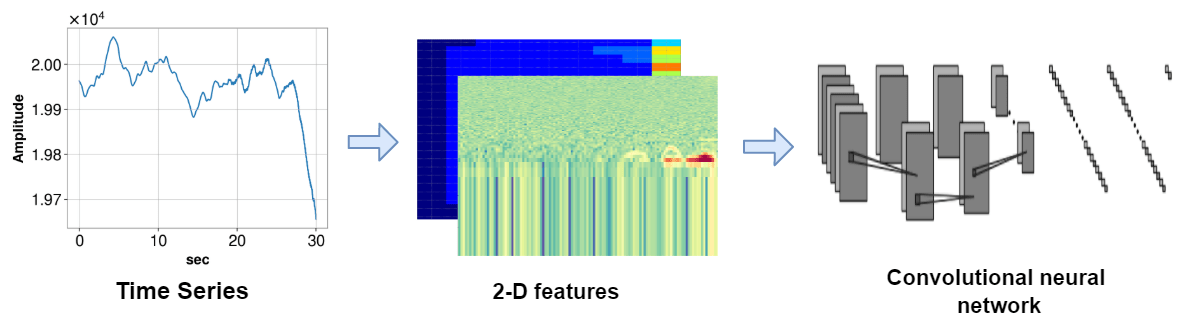}
		\centering
		\caption{Work-flow for CNN classification. The left panel shows a time-series before conversion to a {\it dmdt}-image and spectrogram, as show in the middle panel. These images are fed to CNNs for supervised classification. Once trained, the CNN can then be used in real-time or in archival mode to classify incoming time-series to predict possible lockloss events.}
		\label{fig:cnn}
	\end{figure}

	Random forest (RF; \cite{rf}) is an ensemble learning method where a large number of decision trees are constructed with various subsets of a set of features derived from the input data. The output of RF classification is the mode of the answer given by each decision tree. We used several simple statistical features which have been used for the classification of astronomical light curves~\cite{Richards2011}, as detailed in Table~\ref{table:rf_features}. 
	Chosing a few tens of features has the advantage that each time-series gets reduced from hundreds of thousands of points to just several summary numbers. We note that handcrafting the features is a common practice to improve classification accuracy, but this can also lead to a bias and non-generality. An alternative is to obtain hundreds to thousands of summary statistics, and then use dimensionality reduction techniques to bring the number down for a given problem. This too can introduce some subjectivity. Given that the time-series we have are non-sparse, regular, and continuous, we found using the several standard statistical features described in~\cite{Richards2011} to be sufficient for our targeted problem. Figure~\ref{fig:random forest} shows a schematic of the transformation from a time-series to features to classifications.
	
	\begin{table}
		\centering
		\caption{We list here the time-series features we used with random forest. This  approach was developed for use in optical astronomy~\cite{light_curve}, where it is traditional to use magnitudes (proportional to -1$\times$log(flux)).
		The method is, however, independent of taking --- or not taking --- logarithms. Since dimensionality reduction happens during the workflow, features can be independently computed with or without logarithms. In this paper, we have treated our measured amplitudes like magnitudes, and taken exponentials where amplitudes were expected in the original feature set, thus mapping log-lin values to lin-exp.
	}
		\label{table:rf_features}
\begin{tabular}{lc}
\hline	
Feature  &  Formula  \\ \hline
 mean  &  $<{\rm amplitude}>$   \\
 min  &  ${\rm amplitude}_{{\rm min}}$   \\
 max  &  ${\rm amplitude}_{{\rm max}}$   \\
 halfpeak2peak  &  $0.5*({\rm amplitude}_{{\rm max}}-{\rm amplitude}_{{\rm min}})$   \\
 beyond1std  &  $p(|({\rm amplitude}-<{\rm amplitude}>)|>\sigma)$   \\
 percentile ratio mid20  &  $(e^{\rm amp}_{{\rm 60}}-e^{\rm amp}_{{\rm 40}})/(e^{\rm amp}_{{\rm 95}}-e^{\rm amp}_{{\rm 5}})$   \\
 percentile ratio mid35  &  $(e^{\rm amp}_{{\rm 67.5}}-e^{\rm amp}_{{\rm 32.5}})/(e^{\rm amp}_{{\rm 95}}-e^{\rm amp}_{{\rm 5}})$   \\
 percentile ratio mid50  &  $(e^{\rm amp}_{{\rm 75}}-e^{\rm amp}_{{\rm 25}})/(e^{\rm amp}_{{\rm 95}}-e^{\rm amp}_{{\rm 5}})$   \\
 percentile ratio mid65  &  $(e^{\rm amp}_{{\rm 82.5}}-e^{\rm amp}_{{\rm 17.5}})/(e^{\rm amp}_{{\rm 95}}-e^{\rm amp}_{{\rm 5}})$   \\
 percentile ratio mid80  &  $(e^{\rm amp}_{{\rm 90}}-e^{\rm amp}_{{\rm 10}})/(e^{\rm amp}_{{\rm 95}}-e^{\rm amp}_{{\rm 5}})$   \\
 linear trend  &  b where $ {\rm amplitude}  = a*t + b$   \\
 max slope  &  ${\rm max}(|({\rm amplitude}_{i+1}-{\rm amplitude}_{i})/(t_{i+1}-t_{i})|)$   \\
 median absolute deviation  &  ${\rm med}(e^{\rm amp}-e^{\rm amp}_{{\rm med}})$   \\
 median buffer range percentage  &  $p(|e^{\rm amp}-e^{\rm amp}_{{\rm med}}|<0.1*e^{\rm amp}_{{\rm med}})$   \\
 pair slope trend  &  $p(e^{\rm amp}_{{\rm i+1}}-e^{\rm amp}_{{\rm i}}>0;i=n-30,n)$   \\
 percent difference percentile  &  $(e^{\rm amp}_{{\rm 95}}-e^{\rm amp}_{{\rm 5}}/e^{\rm amp}_{{\rm med}})$   \\ \hline

\end{tabular}
\end{table}
 

\subsubsection{Convolutional neural networks}\label{Convolutional neural networks}
	A convolutional neural network (CNN; \cite{LeCun,deep_learning}) is a supervised method of classification belonging to the deep artificial neural network class of machine learning algorithms. A CNN is an extension of artificial neural network (ANN) \cite{mcculloch1943logical,ANN} and like an ANN it consists of an input layer (typically 2D), an output layer, and multiple hidden layers. The main difference is that it works on smaller parts of an image at a time through kernels and thus there can be translational independence, and at the same time processing of an image can be distributed over several processors of a GPU. The number of layers can also be greater, giving rise to the name {\it deep network}. 
	
	While CNN architectures can be made very complex, we have used layers that are simply stacked. The first couple of layers are purely convolutional and consist of a set of $3x3$ kernels that identify predominant features of that size with successive layers looking for larger features. A maxpool layer reduces the dimensionality by combining adjacent $2x2$ or $3x3$ sets of pixels, the rectified linear units (ReLU) suppresses negative values, while the dropout layer \cite{dropout} helps avoid overfitting by randomly zeroing a fraction of its inputs. The complex interplay of these layers generates thousands to millions of features that are fed to fully connected layers like in an ANN, culminating in a non-linear function to obtain the classifications. The internal weights are then tuned based on known classes from the training data, and a loss-function. This iterative process constitutes the training of the CNN. We analyzed images prepared as detailed in Section~\ref{data_prep} in our analysis.
	
	\subsubsection{t-embedded stochastic neighbour embedding}\label{tsne} 
\begin{figure}
		\includegraphics[width=0.9\textwidth]{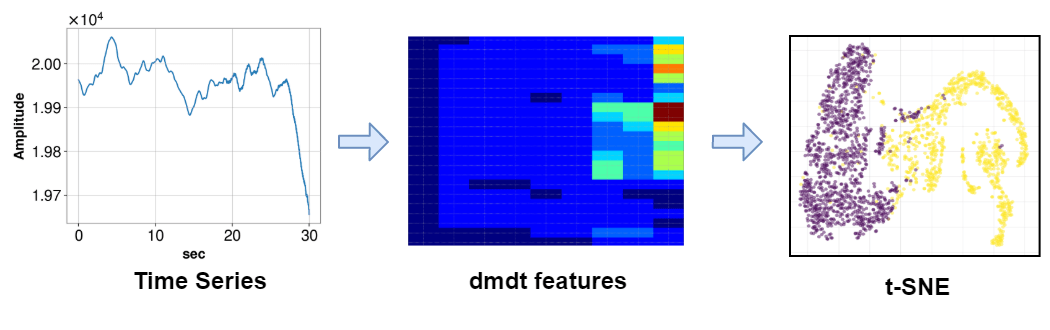}
		\centering
		\caption{Work-flow for t-SNE. Time-series (left panel) are converted to \textit{dmdt}-images (middle panel) and these form input to t-SNE (example in the right panel). While t-SNE is unsupervised, we applied known labels (yellow and purple) to clustered output to investigate the effectiveness of each channel in accurately distinguishing lock losses (yellow) from quiet times (purple).}
		\label{fig:tsne}
	\end{figure}

t-distributed Stochastic Neighbor Embedding (t-SNE; \cite{tsne}) is a non-linear dimensionality reduction technique used for the visualization of high dimensional data into 2 or 3 dimensions.
	In this study, we used t-SNE to identify unsupervised clusters for each considered auxiliary channel given vectorised {\it dmdt} features as input, as shown in Figure~\ref{fig:tsne}. We compared the clustered output to labeled lockloss intervals and normal operation intervals (`quiet' hereafter) to provide an indication of whether a given channel can capture lockloss activity and is able to distinguish it from quiet times.
	If quiet times form a single cluster while the lockloss intervals form two to three clusters, this may indicate different causes for lockloss.
	
	\textit{Perplexity} sets the number of effective nearest neighbors. We experimented with a variety of perplexity values of 2, 5, 10, 20, 60, and 100 for 2000 iterations with a learning rate of 200.  We found that t-SNE was not effective in clustering spectrograms, and we did not pursue that further.
	
	We used the Scikit-learn library \cite{sklearn} for implementing random forest and t-SNE. We used Keras \cite{keras} with a tensorflow \cite{tf} backend for training CNNs.


	\section{Results}\label{results}
	We analyzed LIGO-Hanford lockloss events during the second observing run (O2) for possible causes. We applied three different machine learning methods --- random forests, CNN, and t-SNE --- to pre-processed time-series data of a subset of auxiliary channels, as described in Sections~\ref{sec:data},~\ref{data_prep},~and~\ref{analysismethods}.
	To evaluate our approaches, we calculated commonly used metrics that indicate the effectiveness of ML methods, including: 
	\begin{itemize}
	\item \textit{accuracy}: the fraction of correctly classified objects overall, $(TP+TN)/n$ where $TP$ are true positives, $TN$ are true negatives, and $n$ the total number
	\item \textit{precision}: the ratio of correctly classified objects of a class to the total number of objects classified as belonging to that class, $TP/(TP+FP)$ where $FP$ are false positives 
	\item \textit{recall} (or sensitivity): the ratio of correctly classified objects to the actual number of objects belonging to that class, $TP/(TP+FN)$ where $FN$ are false negatives 
	\item \textit{specificity}: the true negative rate, $TN/(TN+FP)$
	\item \textit{$F_1$ score}: the harmonic mean of precision and recall: $2TP/(2TP+FN+FP)$
	\item \textit{Matthew's correlation coefficient}: $(TP*TN-FP*FN)/\sqrt((TP+FP)(TP+TN)(FP+FN)(TN+FN))$
	\end{itemize}
	While accuracy is a good metric overall in case of balanced classes and comparable classification success, precision is useful when the cost of false positives is high, and recall is a better model when the cost of false negatives is higher. The $F_1$ score provides a balance between precision and recall, and MCC is useful when the negatives are meaningful. 
	
	We also evaluated the computational efficiency of each of the three methods with the goal of moving towards a real-time application of the method. All the experiments were performed on Scientific Linux release 7.5 (Nitrogen). Deep learning models were trained on Tesla P4 with 7.4 GB of RAM.
	
	\subsection{t-SNE plots}\label{t-SNE plots}
\begin{figure}
		\centering
		\subfloat[ ]
		{\includegraphics[width=0.43\textwidth]{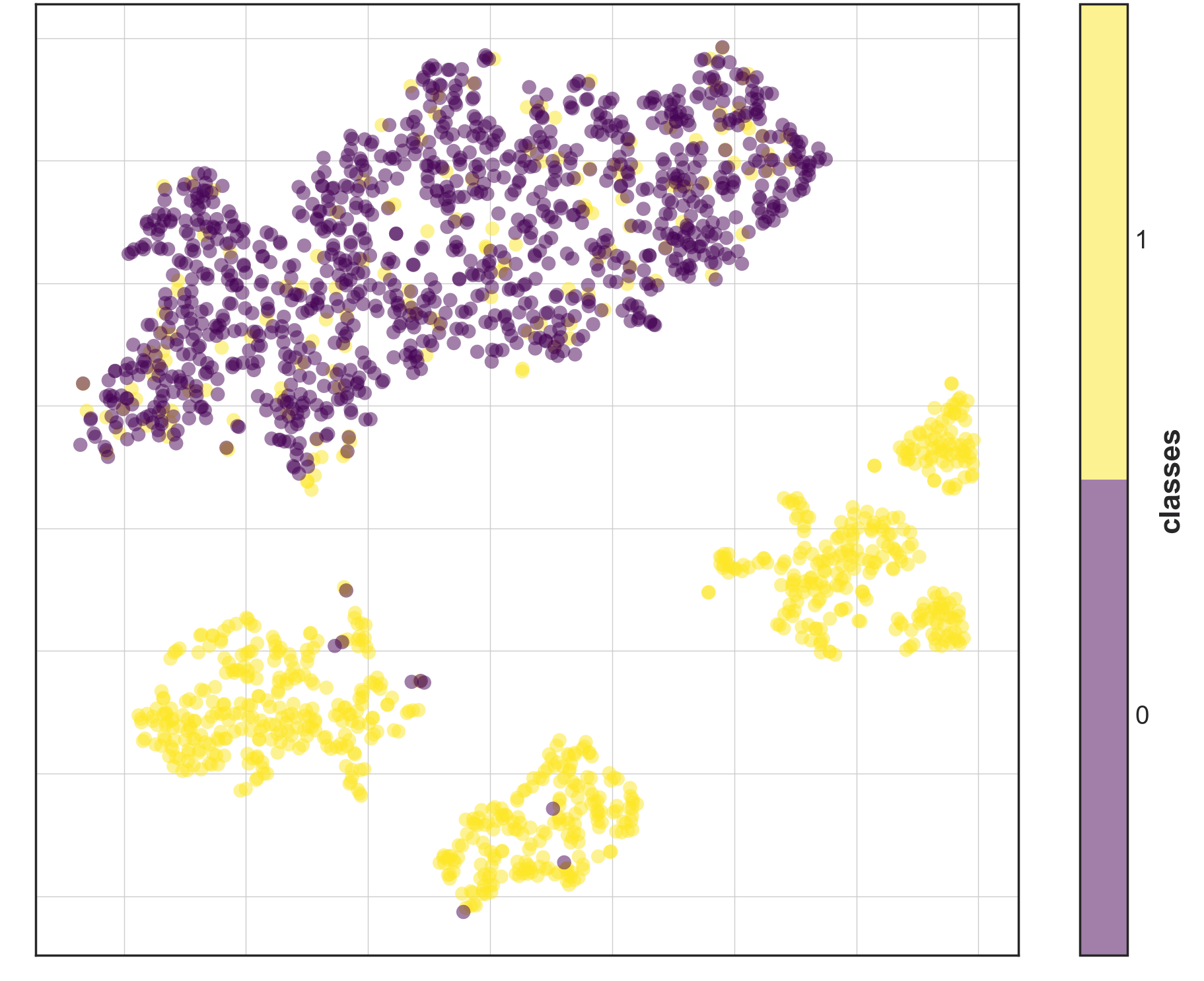}}
		\hfill
		\subfloat[ ]
		{\includegraphics[width=0.43\textwidth]{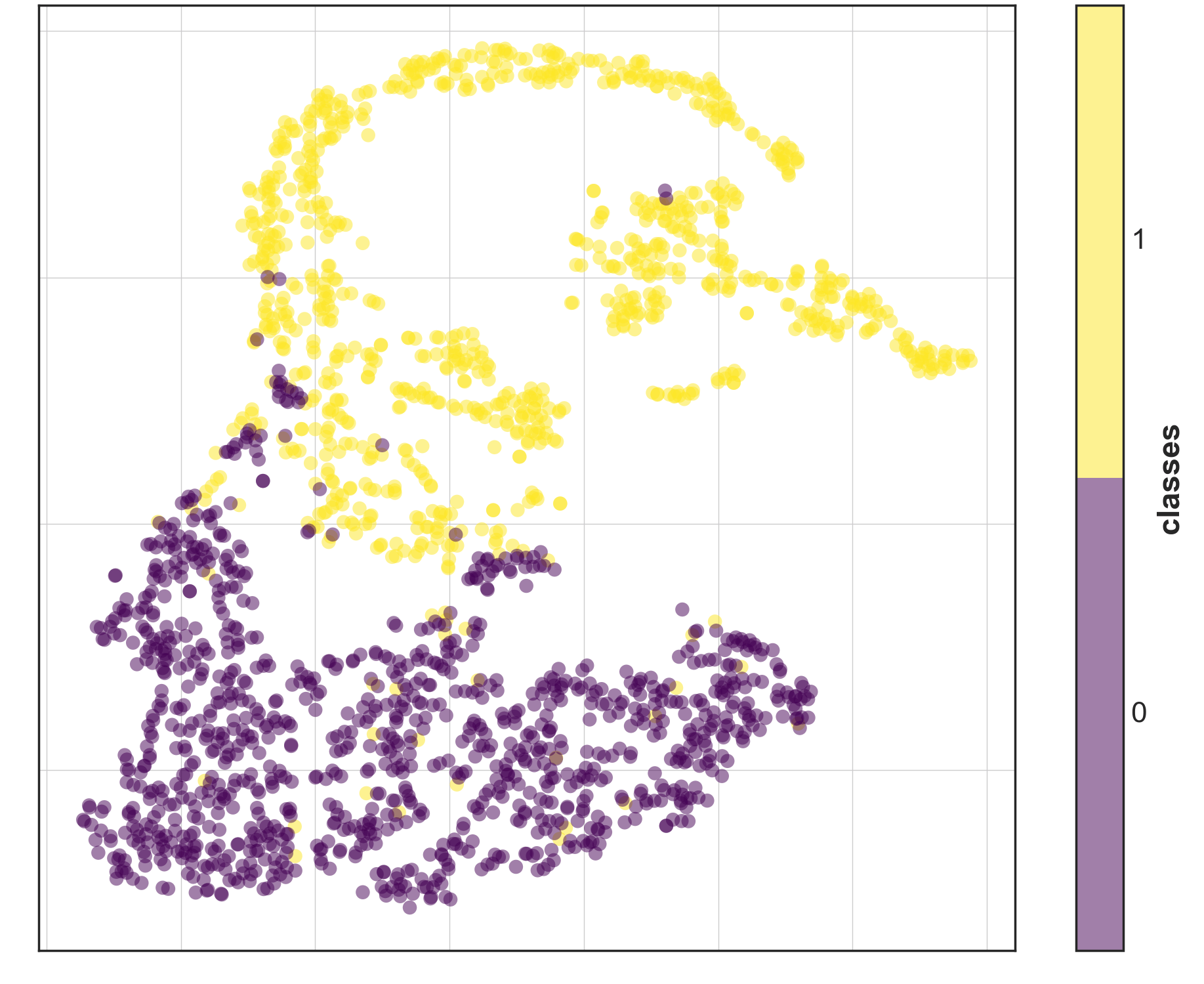}}
		\caption{t-SNE clustering of all LIGO-Hanford lock losses during the second observing run (O2). The left plot shows results using only the SRCL channel and the right plot shows results using a combination of multiple channels (POP+SRCL+MICH). We visualized the accuracy of this unsupervised approach by applying known labels after clustering: yellow points belong to the lockloss class and purple points belong to the quiet class. Note that t-SNE reduces the input dimensions to 2 arbitrary dimensions.  
		}
		\label{fig:cavity_tsne}
	\end{figure}

	We performed t-SNE on the \textit{dmdt} features of single channels as well as on the combined \textit{dmdt} features of multiple channels where we stacked single channels to form a 3D \textit{dmdt} feature. These results indicate how well each channel or channel combination can differentiate between lockloss and quiet times. Compared with the CNN method, discussed later in this section, we observe fewer outliers in the t-SNE plots of combined \textit{dmdt}, as shown in Figure \ref{fig:cavity_tsne}.
	In the future when we characterize the outliers, and implement a real-time version, a quick mapping onto such a t-SNE mapping will be useful to quickly determine the nature of the event.
	
\begin{figure}
		\centering
		\subfloat[ ]
		{\includegraphics[width=0.45\textwidth]{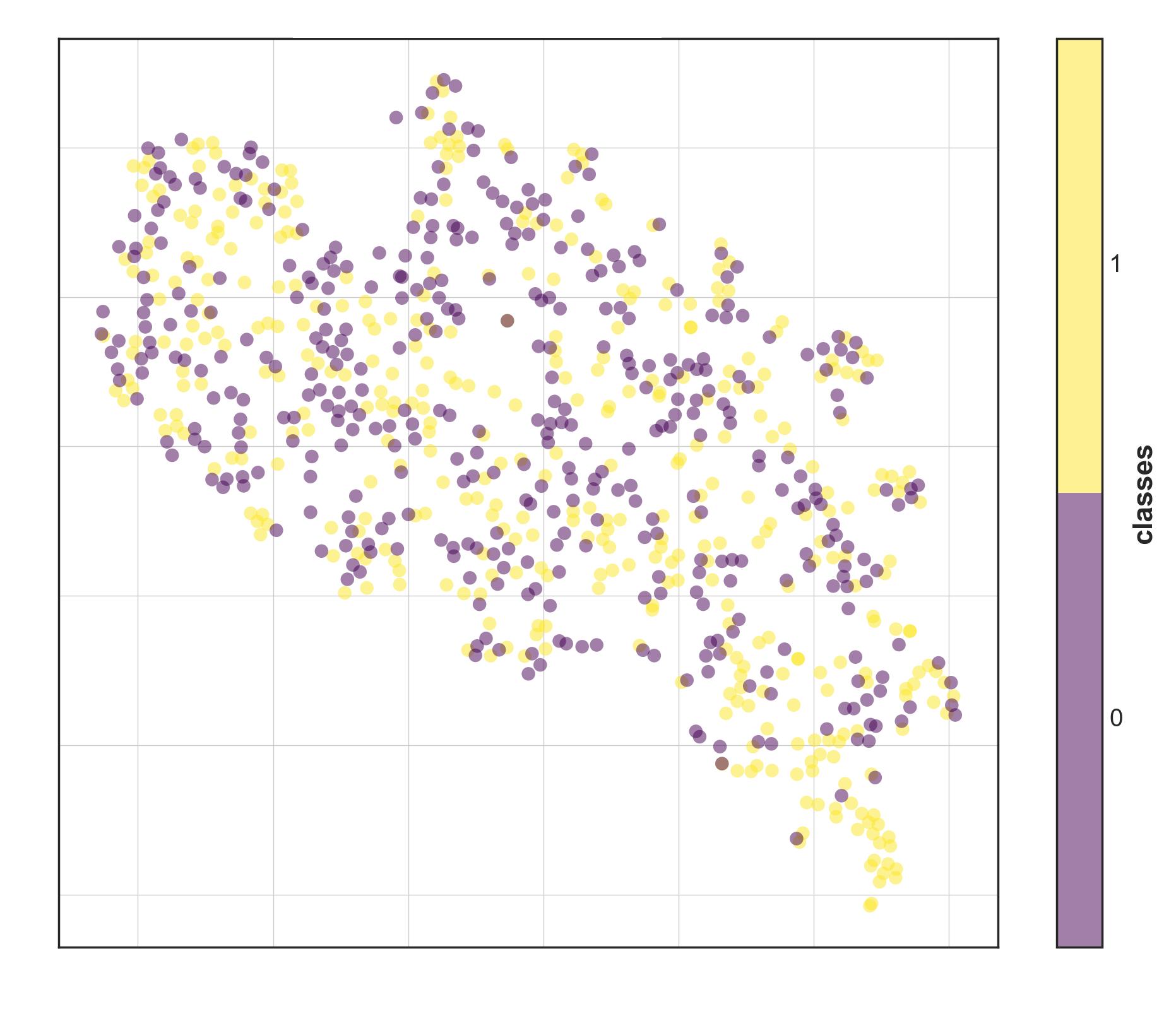}}
		\hfill
		\subfloat[ ] {\includegraphics[width=0.45\textwidth]
			{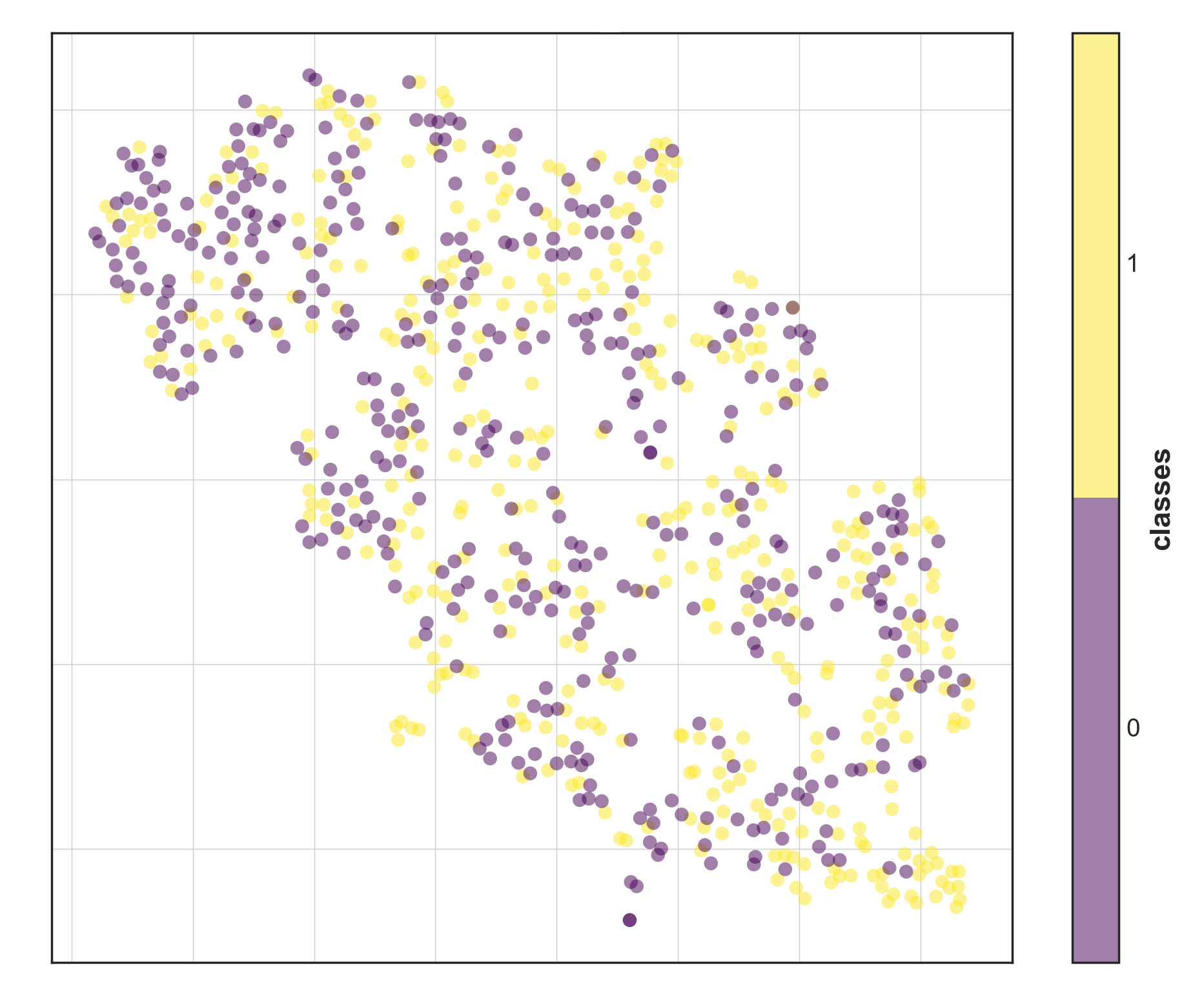}}
		\caption{t-SNE on \textit{dmdt} features of ground motion channels. 0.03-0.1 Hz vertical ground motion, which is useful for capturing earthquake activity, is shown on the left, and 0.03-0.1 Hz horizontal ground motion, useful for capturing wind-induced ground motion, is shown on the right. Here we also applied known labels after clustering: yellow points indicate lockloss times and purple indicate quiet times. Ground motion channels show no evidence of clustering into lockloss and quiet times, unlike the interferometer sensing and control channels shown in Figure \ref{fig:cavity_tsne}.  
		}
		\label{fig:env_tsne}
	\end{figure}

	Notably, when we performed t-SNE on the \textit{dmdt} features of environmental ground motion channels, we observed no well defined clusters, as shown in Figure ~\ref{fig:env_tsne}. We found that none of the features we derived from the ground motion channels we analyzed, described in Section \ref{Auxiliary channel data}, were useful distinguishers between lockloss times and quiet times. This result does not contradict the previously reported correlation, on the scale of tens of minutes or more, between seismic events and lockloss events, but it does give strong evidence that interferometer control channels are a much better indicator of when an interferometer is about to lose lock. 
	
	\subsection{Random forest}\label{random forest}
	
	We performed binary classification using random forests with features described in Table \ref{table:rf_features} computed for each of the interferometer sensing and control channels described in Section \ref{Auxiliary channel data}. With an analysis of individual channels, we obtained the highest accuracy (92.4\%) for the Signal Recycling Cavity Length (SRCL) control channel, as shown in Table \ref{table:Random forest on single channel}. This result implicates this element of the interferometer as indicative of a strong majority of lockloss events.

	\begin{table}
		\centering
		\caption{Random forest on single channels.}
		\label{table:Random forest on single channel}
		\begin{tabular}{lcccccc}
			\hline
			Channel & Test  & Matthew's  & F1 score & Specificity & Recall & Precision \\
			&accuracy&coefficient&&&& \\ \hline
			IMC     & 0.851         & 0.714                 & 0.817    & 0.964       & 0.72   & 0.944     \\
			MICH    & 0.902         & 0.807                 & 0.887    & 0.964       & 0.831  & 0.951     \\
			POP     & 0.902         & 0.806                 & 0.888    & 0.956       & 0.839  & 0.943     \\
			PRCL    & 0.888         & 0.787                 & 0.865    & 0.985       & 0.775  & 0.979     \\
			REFL    & 0.898         & 0.798                 & 0.883    & 0.956       & 0.831  & 0.942    \\
			SRCL    & 0.924         & 0.852                 & 0.911    & 0.989       & 0.847  & 0.985     \\
 \hline

		\end{tabular}
	\end{table}
	
	We also analyzed combinations of channels by implementing a soft majority vote, where the vote is decided by taking the mean of probabilities obtained from single channel classification on the random forest classifiers trained on single channels. As shown in Table \ref{table:Accuracy with soft majority voter on multiple channel}, we obtained a maximum of 95.9\% accuracy for a combination of cavity channels. 
	
	\begin{table}
		\centering
		\caption{Accuracy with soft majority voter on multiple channels.}
		\label{table:Accuracy with soft majority voter on multiple channel}
		\begin{tabular}{lcccccc}
			\hline
			Channel        & Test & Matthew's & F1 score & Specificity & Recall & Precision \\ 
			&accuracy&coefficient&&&& \\ \hline
			IMC+POP+REFL   & 0.912         & 0.827                 & 0.898    & 0.974       & 0.839  & 0.966     \\
			MICH+PRCL+SRCL & 0.929         & 0.865                 & 0.918    & 0.996       & 0.852  & 0.995     \\
			All the above  & 0.959         & 0.919                 & 0.954    & 0.996       & 0.915  & 0.995    \\ \hline
		\end{tabular}
	\end{table}
	
\begin{figure}
		\centering
		\subfloat[]{\includegraphics[width=0.34\textwidth,height=0.36\textwidth]{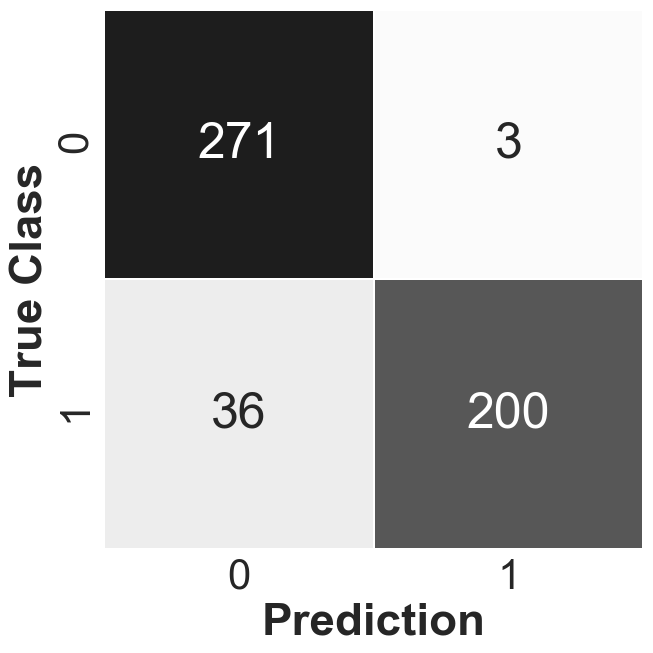}}
		\hfill
		\subfloat[]{\includegraphics[width=0.34\textwidth,height=0.36\textwidth]{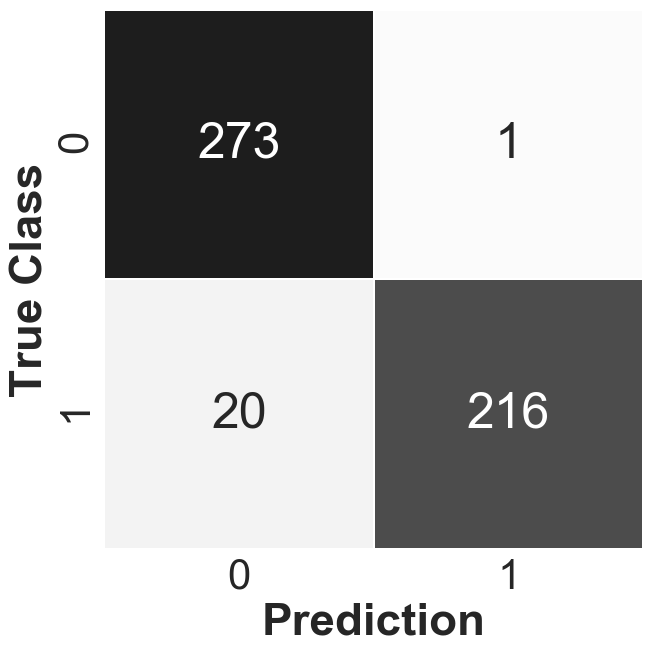}}
		\caption{Confusion matrix for random forest classifications: (a)  for channel SRCL,  (b)  for soft majority voter on cavity channels. We have used label 1 for lockloss events, and label 0 for quiet times. The top-left square indicates the true positives (TP), the bottom-right has true-negatives (TN), the bottom-left includes the false-positives (FP), and the top-right shows the false-negatives (FN). These values are then used to evaluate the metrics like accuracy, recall etc. shown in the different tables as described at the start of Section \ref{results}.
		}
		\label{fig:rf_conf_mat}
	\end{figure}

	Confusion matrices for the SRCL channel, the most accurate single channel predictor, and for a combination of all cavity channels are shown in Figure \ref{fig:rf_conf_mat}. In both cases, true lockloss times are confused for quiet times more often than vice versa by over a factor of 10.
	
	\subsection{CNN with spectrograms}\label{CNN with spectrogram}
	
\begin{figure}
		\includegraphics[width=0.35\textwidth]{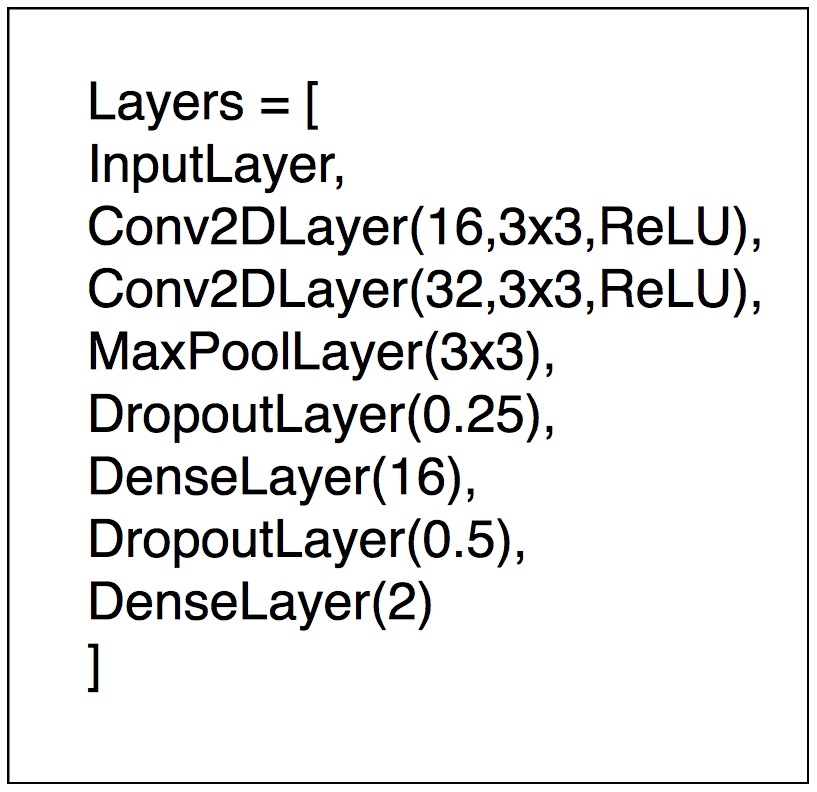}
		\centering
		\caption{Layers of the CNN used for classification of spectrograms. What the layers mean is described in Section. \ref{Convolutional neural networks}. For this CNN configuration and the CNN described in Figure~\ref{fig:dmdt_layers}, after some experimentation we used the L2 regularization with a learning rate of 0.0001, a softmax activation, categorical crossentropy loss function, and the Adadelta optimizer \cite{adadelta}}
		\label{spectrogram_layers}
	\end{figure}

	We performed classification with a CNN, as described in Figure \ref{spectrogram_layers}, on both normalised and unnormalised spectrogram images. We obtain higher accuracy on unnormalised spectrograms than normalised spectrograms. We analyzed sensing and control channels individually, and obtained the highest accuracy (90\%) for the MICH channel, as shown in Table \ref{table:dmdt_spectrogram}. 
	
	We also analyzed combinations of channels by feeding the CNN spectrograms of multiple channels. We obtained a maximum accuracy of 93.5\% with spectrograms of all six cavity channels as input, as reported in Table \ref{CNN accuracy and loss on multiple channel spectrogram}.
	
	\begin{table}
		\centering
		\caption{CNN accuracy and loss for single-channel unnormalised spectrograms.}
		\label{table:dmdt_spectrogram}
		\begin{tabular}{lcccccc}
			\hline
			Channel & Test  & Matthew's & F1 score & Specificity & Recall & Precision \\
			&accuracy&coefficient&&&& \\ \hline
			IMC     & 0.825         & 0.659                 & 0.787    & 0.938       & 0.695  & 0.906     \\
			MICH    & 0.900           & 0.803                 & 0.885    & 0.960        & 0.831  & 0.947     \\
			POP     & 0.853         & 0.717                 & 0.820     & 0.964       & 0.725  & 0.945     \\
			PRCL    & 0.851         & 0.729                 & 0.808    & 1.000           & 0.678  & 1.000         \\

			REFL    & 0.847         & 0.717                 & 0.805    & 0.989       & 0.682  & 0.982   \\
			SRCL    & 0.878         & 0.758                 & 0.861    & 0.934       & 0.814  & 0.914     \\

 \hline 
		\end{tabular}
	\end{table}

	\begin{table}
		\centering
		\caption{CNN accuracy and loss for multiple-channel unnormalised spectrograms.}
		\label{CNN accuracy and loss on multiple channel spectrogram}
		\begin{tabular}{lcccccc}
			\hline
			Channel        & Test & Matthew's  & F1 score & Specificity & Recall & Precision \\
			&accuracy&coefficient&&&& \\ \hline
			IMC+POP+REFL   & 0.845         & 0.71                  & 0.804    & 0.982       & 0.686  & 0.97      \\
			MICH+PRCL+SRCL & 0.888         & 0.79                  & 0.864    & 0.993       & 0.767  & 0.989     \\
			All the above  & 0.935         & 0.875                 & 0.926    & 0.993       & 0.869  & 0.99     \\ \hline
		\end{tabular}
	\end{table}
	
\begin{figure}
		\centering
		
		\subfloat[]{\includegraphics[width=0.34\textwidth,height=0.36\textwidth]{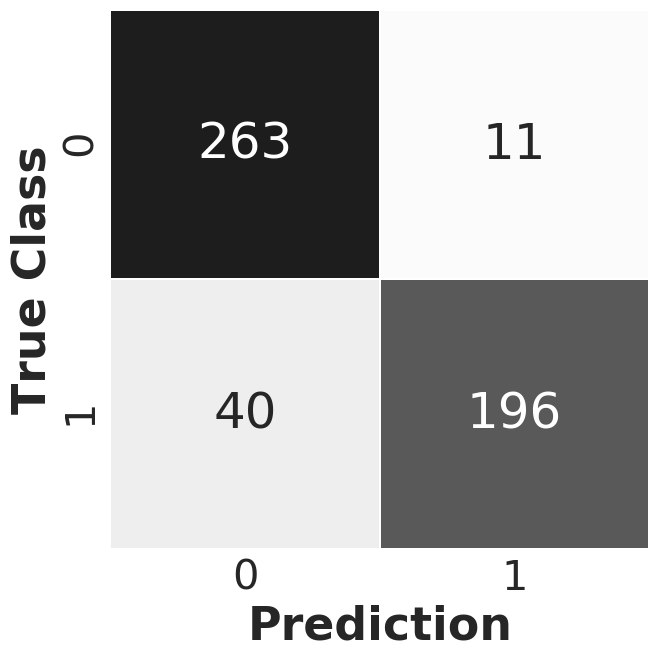}}
		\hfill
		\subfloat[]{\includegraphics[width=0.34\textwidth,height=0.36\textwidth]
			{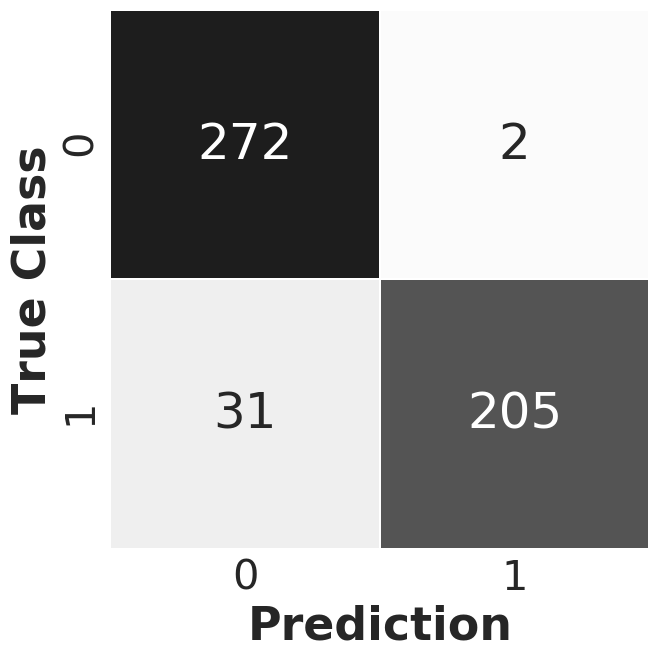}}
		\caption{Confusion matrix for CNN with spectrograms for (a) single channel, MICH; and (b) a combination of cavity channels. Labels : 0 (quiet), 1 (lockloss).}
		\label{fig:spec_conf_mat}
	\end{figure}

	Confusion matrices for the individual MICH channel and the same combination of cavity channels as reported in 
	Table \ref{CNN accuracy and loss on multiple channel spectrogram} are shown in Figure \ref{fig:spec_conf_mat}. As with our random forest results, here again we see lockloss events are more likely to be mistaken for quiet times. 
	
	\subsection{CNN with {\it dmdt}}\label{CNN with dmdt}
	
\begin{figure}
		\includegraphics[width=0.35\textwidth]{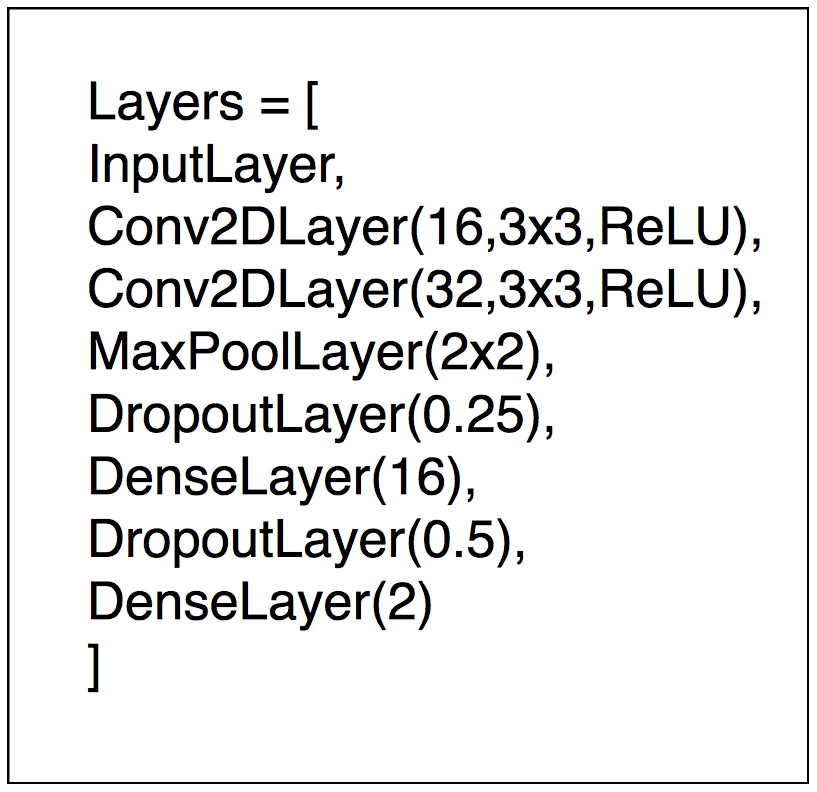}
		\centering
		\caption{Layers of CNN used for classification with {\it dmdt} features.}
		\label{fig:dmdt_layers}
	\end{figure}

	We also performed classification with a CNN, as described in Figure \ref{fig:dmdt_layers}, on \textit{dmdt} features of single channels. We obtained a maximum accuracy of 92.5\% for single channel \textit{dmdt} features with the SRCL channel, as shown in Table \ref{table:CNN accuracy and loss on single channel dmdt}. Again, this result is similar to individual channel analyses that employed random forests and indicates that this single channel can accurately predict most lockloss events.
	
	\begin{table}
		\centering
		\caption{CNN accuracy and loss for {\it dmdt} of single channels.}
		\label{table:CNN accuracy and loss on single channel dmdt}
		\begin{tabular}{lcccccc}
			\hline
			Channel & Test  & Matthew's & F1 score & Specificity & Recall & Precision\\ &accuracy&coefficient&&&&\\ \hline
			IMC     & 0.851         & 0.716                 & 0.816    & 0.971       & 0.712  & 0.955     \\
			MICH    & 0.906         & 0.816                 & 0.890     & 0.974       & 0.826  & 0.965     \\

			POP     & 0.906         & 0.818                 & 0.889    & 0.982       & 0.818  & 0.975     \\
			PRCL    & 0.886         & 0.786                 & 0.861    & 0.993       & 0.763  & 0.989     \\
			REFL    & 0.908         & 0.823                 & 0.891    & 0.985       & 0.818  & 0.980     \\
			SRCL    & 0.925         & 0.857                 & 0.913    & 0.993       & 0.847  & 0.990      \\

 \hline
		\end{tabular}
	\end{table}
	As with our spectrogram CNN analysis, here we also analyzed combinations of cavity channels by stacking \textit{dmdt} features of multiple channels. We obtained the highest accuracy of all of our attempted approaches, 98.6\%, with CNN using stacked \textit{dmdt} features of all six cavity channels reported in Table \ref{CNN accuracy and loss on stacked  dmdt of multiple channel}. 
	
	\begin{table}
		\centering
		\caption{CNN accuracy and loss for stacked {\it dmdt} of multiple channels.}
		\label{CNN accuracy and loss on stacked  dmdt of multiple channel}
		\begin{tabular}{lcccccc}
			\hline
			Channel        & Test  & Matthew's  & F1 score & Specificity & Recall & Precision \\ 
			&accuracy&coefficient&&&& \\ \hline
			IMC+POP+REFL   & 0.913         & 0.832                 & 0.900      & 0.978       & 0.839  & 0.971     \\
			MICH+PRCL+SRCL & 0.968         & 0.938                 & 0.965    & 0.996       & 0.936  & 0.995     \\
			All the above  & 0.986         & 0.973                 & 0.985    & 0.996       & 0.975  & 0.996    \\ 
			\hline
		\end{tabular}
	\end{table}
	
\begin{figure}
		\centering
		\subfloat[]{\includegraphics[width=0.34\textwidth,height=0.36\textwidth]{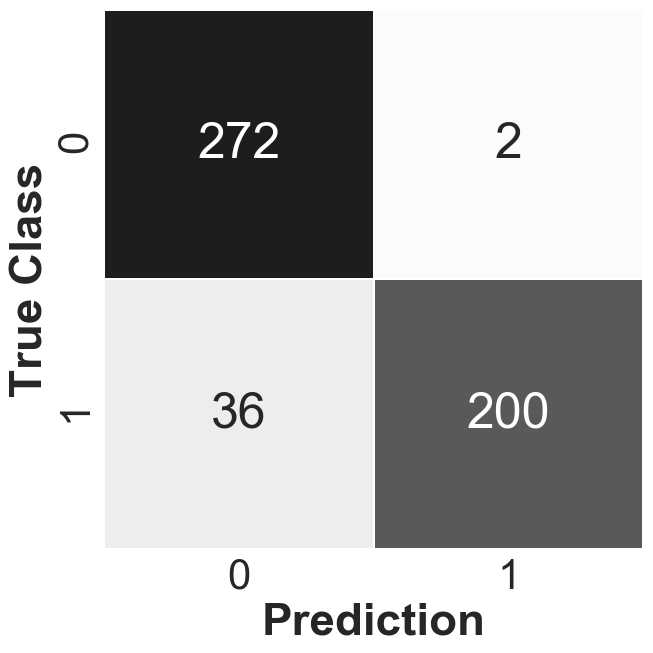}}
		\hfill
		\subfloat[]{\includegraphics[width=0.34\textwidth,height=0.36\textwidth]{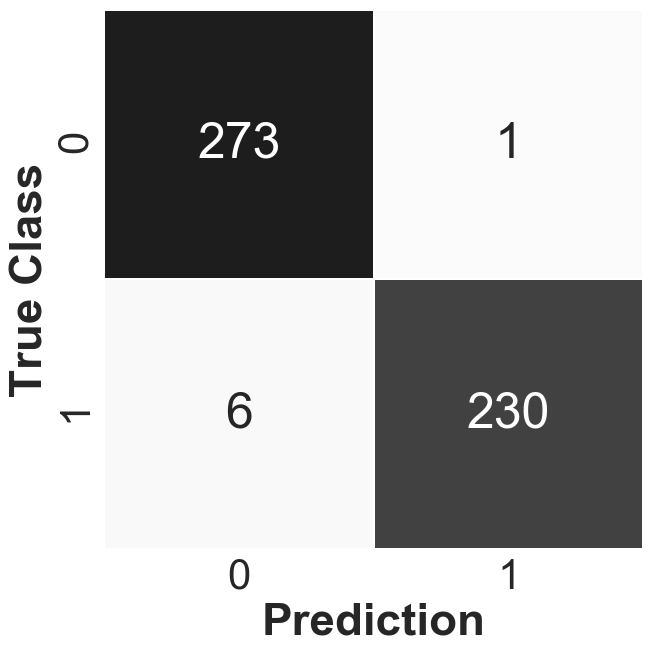}}
		\caption{Confusion matrix for {\it dmdt} classification (a) for single channel, SRCL; and (b) for stacked {\it dmdt} of cavity channel. Labels: 0 (quiet), 1 (lockloss). }
		\label{fig:dmdt_conf_mat}
	\end{figure}

	Confusion matrices for the individual SRCL channel and the same combination of cavity channels as reported in 
	Table \ref{CNN accuracy and loss on stacked  dmdt of multiple channel} are shown in Figure \ref{fig:dmdt_conf_mat}. As with our results for random forests and CNN with spectrograms, here too we see lockloss events are more likely to be mistaken for quiet times. 
	
	\subsection{Obtaining a minimal set of channels for lockloss indicator}\label{Obtaining a minimal set of channels for lockloss indicator}
\begin{figure}
		\centering
		\includegraphics[width=0.5\textwidth]{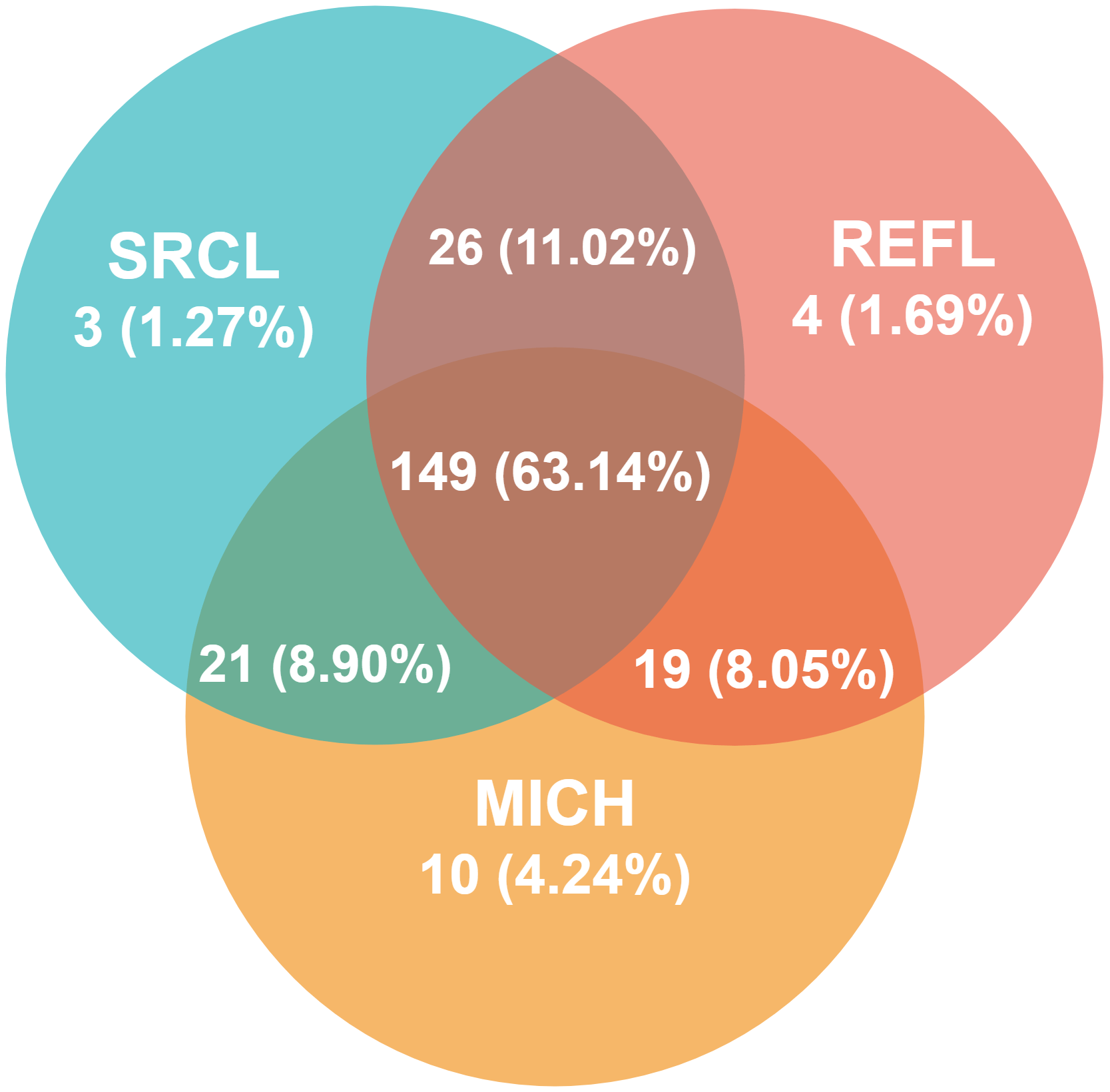}
		
		\caption{Distribution of lockloss captured by the three chosen cavity channels (MICH, REFL, SRCL); Out of 236 test lockloss events, 149 were captured by all three, 66 were captured by at least two of the three channels, and another 17 by at least one of the three, thus leaving only 4 events that could not be captured by any of the three channels.}
		\label{fig:venn_diagram}
	\end{figure}

	We attempted to obtain a small subset of channels which can capture most lockloss activity. For this we derived {\it dmdt} features from all the analyzed channels and trained CNN models on different combinations of stacked {\it dmdt} features. We found that we can achieve an accuracy of 97.5\% with MICH and REFL, 98.6\% with SRCL, REFL and MICH and around the same with combinations of 4 and 5 channels, as shown in Table \ref{table:subset_dmdt}. To visualize the distribution of lockloss events captured by the three best performing channels --- SRCL, REFL and MICH --- we constructed a Venn diagram using the predictions for the test set events obtained from the models trained on each single channel's {\it dmdt} features. We observed that some lockloss events are captured only by a single channel, as shown in Figure \ref{fig:venn_diagram}. Around 63\% of events are captured by all the three channels.
	
	\begin{table}
		\centering
		\caption{Performance metrics for stacked {\it dmdt} of multiple channels.}
		\label{table:subset_dmdt}
		\begin{tabular}{lcccccc}
			\hline
			
			channels               & Test & Matthew's  & F1 score & Specificity & Recall & Precision \\ 
			&accuracy&coefficient&&&& \\ \hline

			MICH+POP               & 0.971         & 0.941                 & 0.968    & 0.985       & 0.953  & 0.983     \\
			MICH+REFL              & 0.975         & 0.949                 & 0.972    & 0.996       & 0.949  & 0.996     \\
			MICH+SRCL              & 0.971         & 0.941                 & 0.968    & 0.989       & 0.949  & 0.987     \\			
			MICH+POP+SRCL          & 0.984         & 0.969                 & 0.983    & 1.000           & 0.966  & 1.000         \\
			MICH+PRCL+REFL         & 0.984         & 0.969                 & 0.983    & 0.996       & 0.970   & 0.996     \\
			MICH+REFL+SRCL         & 0.986         & 0.973                 & 0.985    & 0.996       & 0.975  & 0.996     \\
			IMC+MICH+POP+SCRL     & 0.988         & 0.976                 & 0.987    & 0.996       & 0.979  & 0.996     \\
			IMC+MICH+PRCL+REFL     & 0.986         & 0.973                 & 0.985    & 1.000           & 0.970   & 1.000         \\
			MICH+PRCL+REFL+SRCL    & 0.986         & 0.973                 & 0.985    & 0.996       & 0.975  & 0.996     \\
			IMC+MICH+POP+PRCL+SCRL & 0.990         & 0.980                 & 0.989    & 1.000           & 0.979  & 1.000        
			\\			\hline
		\end{tabular}
	\end{table}

	\subsection{How far in advance can we accurately predict lockloss?}\label{prediction}
	
    Finally, we attempted to predict a lockloss event before its occurrence. 
	We derived {\it dmdt} features from the time-series several seconds before the time of interest and performed binary classification to predict whether lockloss occurs or not. We constructed separate models for prediction of lockloss at 1, 5, 10, 15, 20 and 30 secs prior to lockloss. 
	We observed higher accuracies when we use stacked {\it dmdt} features of multiple channels compared to individual channels. 
	To contrast the performance between single channel and multiple channels, we compared the performance between SRCL (Table \ref{table:predict_single_ch}) and the chosen subset of cavity channels from Figure \ref{fig:venn_diagram} (Table \ref{table:predict_multi_ch}). We observed that in both cases, while the specificity is high for all the times, the predictive accuracy decreases as the time prior to lockloss increases, particularly for single channels. 
	Models trained on stacked {\it dmdt} features performed quite well with a recall above 85\% for time prior to lockloss of less than 15 seconds.
	
	\begin{table}
		\centering
		\caption{Performance metrics for {\it dmdt} images of the SRCL channel.}
		\label{table:predict_single_ch}
		\begin{tabular}{lcccccc}
			\hline
			Time prior  & Test  & Matthew's & F1 score & Specificity & Recall & Precision \\ 
			to lockloss (sec)&accuracy&coefficient &&&& \\ \hline
			0                            & 0.925         & 0.857                 & 0.913    & 0.993       & 0.847  & 0.990      \\
			1                            & 0.884         & 0.779                 & 0.861    & 0.982       & 0.771  & 0.973     \\
			5                            & 0.778         & 0.593                 & 0.695    & 0.978       & 0.547  & 0.956     \\
			10                           & 0.771         & 0.579                 & 0.681    & 0.978       & 0.530   & 0.954     \\
			15                           & 0.767         & 0.570                  & 0.676    & 0.974       & 0.525  & 0.947     \\
			30                           & 0.765         & 0.572                 & 0.669    & 0.982       & 0.513  & 0.960     \\ \hline
		\end{tabular}
		
	\end{table}
	
	\begin{table}
		\centering
		\caption{Performance metrics for stacked {\it dmdt} of SRCL+MICH+REFL.}
		\label{table:predict_multi_ch}
		\begin{tabular}{lcccccc}
			\hline
			Time prior  & Test  & Matthew's & F1 score & Specificity & Recall & Precision \\ 
			to lockloss (sec)&accuracy&coefficient &&&& \\ \hline
			0                            & 0.986         & 0.973                 & 0.985    & 0.996       & 0.975  & 0.996     \\
			1                            & 0.971         & 0.941                 & 0.968    & 0.974       & 0.966  & 0.970      \\
			5                            & 0.927         & 0.860                  & 0.916    & 0.993       & 0.852  & 0.99      \\
			10                           & 0.916         & 0.833                 & 0.904    & 0.964       & 0.860   & 0.953     \\
			15                           & 0.916         & 0.834                 & 0.903    & 0.971       & 0.852  & 0.962     \\
			30                           & 0.900           & 0.806                 & 0.882    & 0.978       & 0.809  & 0.970      \\ \hline
		\end{tabular}
	\end{table}
	
	\section{Summary and discussion}\label{discussion}
	In this work, we outlined a method to investigate which components within interferometric gravitational-wave detectors are most susceptible to a lockloss using machine learning techniques. 
	We investigated auxiliary channels that formed distinct clusters using the t-SNE approach that accurately distinguished between known quiet and lockloss times. 
	We experimented with different features such as statistical features used with RF, as well as {\it dmdt} and spectrograms used with CNNs. We found that classification using RF reaches an accuracy comparable to CNNs in case of single channel classification. However, classification on stacked {\it dmdt} features using CNNs outperformed RF in case of multiple channel classification. 
	
	In an effort to diagnose the interferometer components associated with locklosses, we identified a minimal subset of channels which captured a majority of lockloss events. 
	We found that a minimal set of three cavity channels (SRCL, REFL and MICH) accurately distinguished between lockloss and quiet times for 98\% of the test set examples (232 out of 236 lockless events).
    We also investigated the performance of our method using a set of channels used to sense and control the alignment of optical cavities within the LIGO detectors, and found that these channels were not as good predictors of lockloss, using feature sets as described in Section~\ref{analysismethods}. We leave improvement of feature sets for alignment channels to future study. 
	
	We next investigated the power of this approach in predicting a lockloss before it occurs. Using {\it dmdt} features with a chosen subset of cavity channels, we found the accuracy of this method to be above 90\% for times up to 30 seconds prior to lockloss. This result shows extreme promise for applying this method in near-real-time to accurately predict lockloss events before they occur, which may allow automated systems like the Guardian~\cite{Rollins2016} to change the interferometer configuration to avoid a lockloss. 
	
	As future work, we intend to improve feature extraction and the overall performance of our method by further analyzing commonalities between mis-classified events.
	A further goal is to incorporate a streaming lockloss prediction pipeline for interferometer state changes, complementary to the method described in \cite{Coughlin}, to increase the duty cycle of the global interferometer network.
	As the behavior of the detectors can change quite significantly between observing runs due to instrumentation changes and commissioning work, such a pipeline would need to be re-trained to reflect these differences ahead of deployment. 
	Another natural extension of this work is to target the diagnosis of data quality artifacts in the data to increase the sensitivity of the transient astrophysical searches.
	
	\section*{Acknowledgements}
	The authors thank Jameson Rollins for useful discussion on other relevant work ongoing within the LIGO Scientific Collaboration as well as developing and maintaining the code used to identify precise lockloss times we used for a labelled data set. We also thank Sheila Dwyer for guidance on auxiliary witnesses that would be fruitful to target for this study. 
	The authors are grateful for computational resources provided by the LIGO Laboratory and supported by National Science Foundation Grants PHY-0757058 and PHY-0823459. 
	AB would like to thank IIT Gandhinagar, the Caltech SURF program, and LIGO, Caltech for support during the study.
	AM acknowledges support from the NSF (1640818, AST-1815034). 
	AM and JM also acknowledge support from IUSSTF (JC-001/2017).
	LIGO was constructed by the California Institute of Technology and Massachusetts Institute of Technology with funding from the National Science Foundation, and operates under cooperative agreement PHY-0757058. Advanced LIGO was built under award PHY-0823459. This paper carries LIGO Document Number LIGO-P1900222.
	
	\section*{References}
	
	\bibliographystyle{unsrt}
	\bibliography{ref}


\end{document}